\documentclass[epj,nopacs]{svjour}
\usepackage{graphics}
\usepackage{tikz-feynman}
\usepackage{multirow}
\usepackage{cuted}
\usepackage{float}
\usepackage{amsmath}
\usepackage[numbers, square, sort&compress]{natbib}
\allowdisplaybreaks
\usepackage{etoolbox}
\AfterEndEnvironment{strip}{\leavevmode}

\makeatletter

\def\makeheadbox{\relax}
\def\@makeheadbox{\relax}

\def\ps@headings{\let\@mkboth\markboth
  \def\@oddfoot{}
  \def\@evenfoot{}
  \def\@oddhead{}
  \def\@evenhead{}}
\def\ps@titlepage{\let\@mkboth\markboth
  \def\@oddfoot{}
  \def\@evenfoot{}
  \def\@oddhead{}
  \def\@evenhead{}}
\makeatother

\journalname{Eur. Phys. J. C}
\usepackage[breaklinks=true]{hyperref}
\usepackage{xurl}
\usepackage[table]{xcolor}
\begin{document}

\sloppy 

\title{Semileptonic $D_{e4}$ decays: hadronic dynamics and the determination of $|V_{cs}|$}

\author{J. L. Guti\'errez Santiago\thanks{jorge.gutierrez@cinvestav.mx} and G. L\'opez Castro\thanks{gabriel.lopez@cinvestav.mx}}
%
%
\institute{Departamento de F\'isica, Centro de Investigaci\'on y de Estudios Avanzados,\\ Apdo. Postal 14-740, 07000 Ciudad de M\'exico, M\'exico. 
}
%
%
\abstract{
The four-body decays $D^+ \to K^-\pi^+e^+\nu_e$ ($D_{e4}^+$) and $D^0\to \overline{K^0}\pi^-e^+\nu_e$  ($D^0_{e4}$) are studied  in a model where the momentum-dependence of the hadronic matrix elements are described in terms of $K^*(892)$ and $D^*(2010)$ pole contributions. From fits to the recent data of the BESIII collaboration we find that the $D^*$-pole can mimic the effect of the $S$-wave contribution  of the $K\pi$ system to the branching fraction. Implications for the determination of the $|V_{cs}|$ quark mixing matrix element are discussed. 
}
%
\titlerunning{Semileptonic $D_{e4}$ decays}
\date{}
\maketitle

\section{Introduction}

The precise determination of  the Cabibbo-Kobayashi-Maskawa  (CKM) matrix  elements\citep{Cabibbo:1963, Kobayashi:1973} is very important to test the Standard Model at the quantum level.  Deviations from its unitarity properties ($\sum_{j=d,s,b}V_{ij}V^*_{kj}= \sum_{m=u,c,t}V_{mi}V^*_{mk}=\delta_{ik}$) can provide indications of New Physics. In order to achieve these goals one needs, in addition to very precise data on semileptonic decays, a careful account of radiative corrections and of strong interactions effects. Currently, the most precise test of unitarity involves entries of the first row of the CKM  matrix. This is possible due to detailed calculations of electroweak radiative corrections and a good control of  hadronic form factors uncertainties. From a comparison of precise data with theoretical predictions of similar accuracy one gets $|V_{ud}|^2+|V_{us}|^2+|V_{ub}|^2=0.9985(7)$ \cite{PDG}, which currently exhibits a 2$\sigma$ tension with three-generations unitarity. 
Similarly, the validation of orthogonality of different rows and columns, allow to test the CKM paradigm of Charge-Parity (CP) violation. Most precise tests of orthogonality relations at flavor factories like Belle II \cite{Kou:2018nap, 10.1007/978-3-030-29622-3_38}, 
Belle and Belle II \cite{Belle&Belle-II, ByBII7, johnson2025p}, LHCb \cite{10.1007/JHEP12-2021-141, 10.1007/JHEP12-2023-013, 10.1007/JHEP07(2023)138, 10.1007/JHEP02-2024-118, 10.1140/epjc/s10052-023-12376-z, mackay2024gamma, hao2024ckmgammameasurementslhcb}, and BESIII \cite{Ablikim_2020, PhysRevD.101.112002, zhou2024inputsgammameasurementsbesiii} 
may eventually provide indications of new sources of CP violation. 

Unitarity tests for other relations of the CKM matrix are not equally precise. For instance, using the world  average value of each $|V_{ci}|$ matrix element \cite{PDG} one gets for the second row  $|V_{cd}|^2+|V_{cs}|^2+|V_{cb}|^2=1.001(12)$, with the uncertainty fully dominated by the one from the $|V_{cs}|$ entry. The dominant contribution to this unitarity relation is the $|V_{cs}|$ matrix element, therefore its value needs to be determined with better precision.  Currently, $|V_{cs}|$ is extracted from the leptonic $D_s \to \mu \nu, \tau \nu$ ($D_{s \ell 2}$) and the semileptonic $D \to K \ell \nu$ ($D_{\ell 3}$) decay processes. As in the case of $|V_{ud}|$ and $|V_{us}|$ determinations, using as many sources as possible to extract $|V_{cs}|$ can be useful to improve its precision and to test the consistency of hadronic inputs. In this paper we focus on the four-body $D\to K\pi \ell \nu$ ($D_{\ell 4}$) decay processes, which can provide an independent determination of $|V_{cs}|$ and also can offer an interesting place to study the rich dynamics of the strong interaction in the weak vertex\footnote{For an inclusive determination of $|V_{cs}|$ see for example  Ref. \cite{LatticeVcsInclusive}}. 

The current world average of $D_{\ell 4}$ branching fractions reported by the Particle Data Group \cite{PDG} are: 
\begin{eqnarray} \label{ExpBR}
B(D^+\to K^-\pi^+ e^+\nu_e) &=& (4.02\pm 0.18)\% \ , \nonumber \\
B(D^+\to K^-\pi^+ \mu^+\nu_{\mu}) &=& (3.65\pm 0.34)\% \ ,  \\
B(D^0\to \overline{K^0}\pi^- e^+\nu_e) &=& (1.44\pm 0.03)\%\  . \nonumber 
\end{eqnarray}
The uncertainties on electronic channels are around 3 to 5 \% . This may be improved in the future and provide a determination of $V_{cs}$ at the subpercent level with further improvements on theoretical inputs.

The strong interaction effects are encoded in the decay constants and (momentum-dependent) weak form factors which define the hadronic matrix elements. While $D_s \to \mu \nu$ is described in terms  of a single decay constant $f_{D_s}$,  the semileptonic $D \to K \ell \nu$ decays need two form factors. The hadronic matrix element of the $D_{\ell 4}$ decays, can be described in terms of four form factors\footnote{Only one (three) form factor is numerically relevant for $D_{e3}$ ($D_{e4}$) decay. More form factors can contribute if effective interactions beyond the SM are present (see for example \cite{Shi:2020rkz}).} (see Section 2). The form factors for different $D_{\ell 4}$ and $B_{\ell 4}$ semileptonic decays have been calculated within different theoretical approaches using Lattice QCD, Light-Cone Sum Rules, dispersive methods applied to energy expansion of the form factors, unitarized chiral perturbation theory and dispersion relations, SU(3) symmetry, etc. (for recent literature see \cite{Wang:2022fbk, Cheng:2025fux, Boer:2016iez, Shi:2020rkz, Leskovec:2025gsw, Cheng:2025hxe, Herren:2025cwv}). Some of these calculations are less model-dependent, like the use of SU(3) symmetry to relate different final states, however they lack the necessary precision to match the experimental accuracy reached in $D_{e4}$ decays. In the present paper (see below) we consider a model where, in addition to the dominant contribution of the $K^*(892)$ resonance in the $K\pi$ system, we also include a novel $D^*$ intermediate meson contribution.  

 The $D\to K\pi\ell \nu$ decays are also particularly interesting to study the dynamics of the  $K\pi$ system. Previous analysis (see for instance \cite{KappaBES, GuoKappa}) indicate that the $P$-wave configuration of the $K\pi$ production is dominated by the $K^*(892)$ vector-meson  resonance. A small $S$-wave contribution, attributed to scalar strange resonances or to a non-resonant background, can also be added in the analysis \footnote{It is also reported  in those references that contributions of higher vector and scalar excitations  as well as the $D$-wave configuration of the $K\pi$ system are negligible small.}.    Although in the past some studies did not included this light scalar meson, it is worth nothing that most analysis including chiral symmetry constrains at low energy find a light $K^*_0(700)$ scalar meson  (see for example the review in \cite{PDG} and references therein).  
Different experimental analysis of $D_{e 4}$ decays  \cite{BESIII1,BABAR,BESIIID0bef, BESIIID0} point to a contribution of around $(5\sim 6)$\% to the branching fraction of $D_{e4}$ decays due to the $S$-wave configuration.

In the present work we reanalyze the semileptonic decays $D^+ \to K^{-}\pi^+ e^{+}\nu_{e}$ ($D^+_{e4}$) and $D^0 \to\overline{K^0}\pi^- e^{+}\nu_{e}$ ($D^0_{e4}$) using the data reported recently by the BESIII collaboration \cite{BESIII1, BESIIID0}.
We assume that the $P$-wave configuration of the $K\pi$ system is dominated by the $K^*(892)$ vector meson resonance.   As it was proposed in Refs. \cite{Kim:2016yth, SGK}, we also consider  a novel contribution of the $D^*$ meson pole which is shown in Figure \ref{FeynmanDiag}(b). A theoretical estimate of the latter effects in the branching fractions of $D_{e4}$ and $B_{\ell 4}$ decays was given already in Ref. \cite{Kim:2016yth, SGK}. It was found in those references that the two-meson mass distributions in four-body decays are affected by the new ($D^*, B^*$) pole contribution. Since this can affect the relative contributions of $P$- and $S$-wave, as well as the determination of the weak form factors and of the $|V_{cs}|$ matrix element, this will be the focus of our present study.  Overall, we find that the $D^*$ pole contribution can mimic the effect of the $S$-wave contribution in the branching fraction of $D_{e4}$ decays.

In order to have a closer  comparison with the analysis of the BESIII collaboration \cite{BESIII1}, in this paper we use the same form factors for the weak $D\to K^*$ transition and we attribute the effect of the $S$-wave contribution to the interference of the $K^*$ and $D^*$-pole amplitudes in our model. Using a different resonant shape for the $p$-wave contribution or a different parameterization of the weak form factors, would not allow to compare ours and BESIII results. 

 We organize our paper as follows: In section \ref{DAK} we discuss the decay and comment briefly on the kinematics of four body decays.  In section \ref{FF} we consider  the form factors required to describe the hadronic currents and discuss the parametrizations including the novel contribution of the $D^*$ meson pole. In subsection \ref{sb}, we compare the  Breit-Wigner parametrizations used in this paper and in the analysis of the  BESIII collaboration. Some technical issues related to the fit procedure used in this paper are discussed in Section \ref{FitCriteria}.
In section \ref{RyD} we present and discuss our results. Finally, in section \ref{Conclusions} we present our conclusions.  We describe in the appendices some details regarding the four-body kinematics and the squared amplitude.

\section{Decay Amplitude and Kinematics}\label{DAK}

\begin{figure*}
\centering
\includegraphics{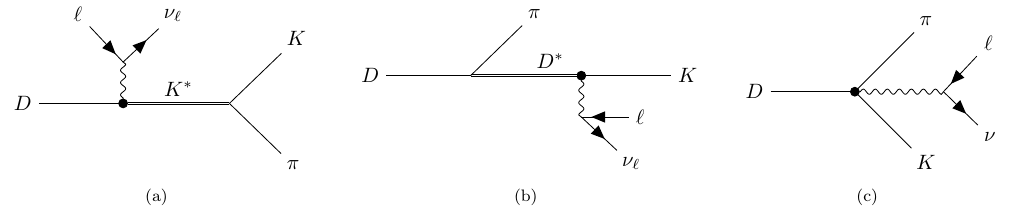}
  \caption{Feynman diagrams for $D_{\ell 4}$ decays: (a) $K^*$ (vector and scalar) meson resonance dominance of the $K\pi$ system  
(b) $D^*$ pole contribution and, (c)  contact term. The hadronic weak vertex is represented by a solid dot. }
  \label{FeynmanDiag}
\end{figure*}

We use the momentum convention $D\left(p \right) \to K\left(P_{1}\right) \pi\left( P_{2}\right) \ell^{+}\left(P_{3}\right) \nu_{\ell}\left(P_{4}\right)$ for a generic $D_{\ell 4}$ decay. The mass of the decaying particle is denoted by $M$ and $m_i$ ($i=1, 2,3,4$) are used for the masses of the final-state particles, respectively. At the lowest order, the amplitude can be factorized in the following form 

\begin{equation}\label{amplitude}
	\mathcal{M} = \frac{G_{F}}{\sqrt{2}}V_{cs} h_{\mu} \ell^{\mu},
\end{equation}

\noindent where $V_{cs}$ is the corresponding element of the CKM matrix, $\ell^{\mu}=\overline{u} (P_{4}) \gamma ^{\mu} (1 - \gamma_{5}) v(P_{3})$ is the leptonic weak current and $h_{\mu} = V_{\mu} - A_{\mu}$ the hadronic matrix element of the weak current.

It becomes convenient to define the following combination of momenta	
\begin{eqnarray}
P &=& P_{1} + P_{2}, \  Q = P_{1} - P_{2}, \nonumber \\   L &=& P_{3} + P_{4}, \  N = P_{3} - P_{4}\ . 
\end{eqnarray}
The most general parametrization of the hadronic weak current depends on four dimensionless form factors \cite{Cabibbo, PhysRev.171.1457, Pais-Treiman, Rosselet:1976pu, Bijnens:1994ie}
\begin{eqnarray} \label{most-general-decomposition}
\!\! h_{\mu} &=& \langle K\pi |\bar{s}\gamma_{\mu}(1-\gamma_5)c| D \rangle \nonumber \\
&=&\! - \frac{H}{M^3}\epsilon_{\mu\nu\rho\sigma} L^{\nu} P^{\rho} Q^{\sigma} + \frac{i}{M}\!\left[\!\frac{}{} F P_{\mu} + G Q_{\mu} + R L_{\mu} \right], \nonumber \\ 
\end{eqnarray}
\noindent namely, one vector ($H$) and three axial vectors ($F, G, R$) form factors, which in general depend on three kinematical Lorentz scalar variables. 

Within the model under consideration illustrated in Figure \ref{FeynmanDiag}, each form factor receives three contributions: from the $K^*$ (a) and $D^*$ (b) poles and from the contact term (c). Within the meson-dominance model, we will assume that the contact term has a negligible contribution (as is done also in other analysis). While the $K^*$ diagrams in Figure 1(a) can receive contributions from the $P$- and $S$-waves according to the spin of the intermediate resonance, the $D^*$ pole can, in principle, contribute to both wave configurations. In this paper we will consider only the lowest mass vector $D^*$ pole contributions; since it is the closest to the physical region, one may expect that higher mass excitations will be further suppressed.

A four-body decay can be described in terms of five independent kinematical variables. We will choose them as follows: ($s_{12}$, $s_{34}$, $\theta_{12}$, $\theta_{34}$ and $\phi$) \cite{Cabibbo}. They are illustrated in Figure \ref{KinFig} and their definition and kinematical limits are given in Appendix \ref{Appendix A}.

\begin{figure}
		\begin{tikzpicture}
			\draw[gray!9, ultra thick, fill=gray!9] (-3.0,0.2) -- (3.4,0.2) -- (3.4,4.2) -- (-3.0,4.2) -- cycle;
			\draw[gray!9, ultra thick,fill=gray!9] (3.1,-1.05) -- (3.1,-5.0) -- (-2.7,-2.6) -- (-2.7,1.35) -- cycle;
                \draw[blue!18, ultra thick,fill=blue!18] (0,0) -- (2.9,-1.2) -- (2.9,-5.2) -- (-2.9,-2.8) -- (-2.9,1.2) -- cycle;
			\draw[green!18, ultra thick, fill=green!20] (-3.2,0) -- (0,0) -- (3.2,0) -- (3.2,4) -- (-3.2,4) -- cycle;
                \draw[blue!20, thick, dashed] (-2.9,0) -- (-2.9,1.2);
                \draw[blue!20, thick, dashed] (-2.9,1.2) -- (0,0);
                \draw[gray, thick, dashed] (0,4.0) -- (0,-4.0);
                \draw[black, thick, ->] (0,2.0) -- (0,3.5) node [above left, pos =-0.01] {$K^{*}$} ; 
                \draw[black, thick, ->] (0,2.0) -- (-1.2,0.75) node [below right, pos =1.0] {$\pi$}; 
                \draw[black, thick, ->] (0,2.0) -- (1.2,3.25) node [above, pos =1.0] {$K$};	
                \draw[black, thick, ->] (0,-2.0) -- (0,-3.3) node [below right, pos =-0.01] {$W$} ;
                \draw[black, thick, -] (0,2.0)--(0,-2.0);
                \draw[black, thick, ->] (0,-2.0) -- (1.5,-1.4) node [above, pos =1.0] {$\nu_{\ell}$};
                \draw[black, thick, ->] (0,-2.0) -- (-1.5,-2.6) node [below, pos =1.0] {$\ell$}; 
                \draw[gray, thick, dashed] (0,-0.03) -- (3.5,-0.03);
			\draw[black, thick, ->] (1.0,-0.4) arc (315:351:0.7) node [right,pos=0.3] {$\phi$};
                \draw[black, thick, ->] (0,2.85) arc[start angle=90, end angle=45, radius=0.8] node [above, pos=0.6]{$\theta_{12}$};
			\draw[black, thick, ->] (0,-2.85) arc(270:200:0.8) node [below,pos=0.5] {$\theta_{34}$};
            \draw[gray, thick, dashed] (-3.3,-2.6) -- (0.03,0);
		\end{tikzpicture}
		\caption{Kinematics of the four-body  $D\to K\pi \ell\nu$ decay (see for example  \cite{Cabibbo}).}\label{KinFig}
\end{figure}

 In terms of these independent variables, the differential decay rate gets the following expression (the definition of different  factors are provided in Appendix \ref{Appendix A})
\begin{equation}
	d \Gamma = \frac{X \beta_{12} \beta_{34}}{4(4\pi)^6 M^3} \overline{\left | \mathcal{M} \right | ^2} \sin \theta_{12} \sin \theta_{34}{\rm d}s_{12} {\rm d}s_{34} {\rm d} \theta_{12} {\rm d} \theta_{34} {\rm d} \phi  \ . 
\end{equation}

 The averaged and unpolarized squared amplitude $\overline{|\mathcal{M}|^2}$ becomes a function of the set of independent variables.  Following a similar kinematical analysis for $K_{e4}$ kaon semileptonic decays \cite{Pais-Treiman, Semi-Bijnens}, it will become convenient to expand $\overline{|\mathcal{M}|^2}$ in terms of the angles $\theta_{34}$ and $\phi$, to leave the final integration over the $s_{12}$, $s_{34}$ and $\theta_{12}$ variables which depend basically on momenta associated to the hadronic vertex.  The expression for this expansion is deserved to  Appendix \ref{Appendix B}.

\section{Hadronic matrix element}\label{FF}

\subsection{Form factors}
In the model under consideration in this paper, the hadronic matrix element for the $D\to K \pi \ell^+ \nu_{\ell}$ decay is given by the following expresion (each term can be easily identified  in Figure \ref{FeynmanDiag})
\vspace{0.05cm}
\begin{strip}
\begin{eqnarray}\label{sumofcontributions}
h_{\mu}=\langle K\pi | j_{\mu}| D\rangle = \sum_{K^*} \frac{\langle K \pi | K^{*} \rangle \langle K^* | j_{\mu}| D\rangle}{D_{K^*}(s_{12})} + \sum_{D^*}\frac{\langle K | j_{\mu} | D^{*} \rangle \langle D^* \pi | D\rangle}{D_{D^{*}}(s')} + C\ . 
\end{eqnarray}
\end{strip}
\vspace{0.05cm}
Here, $j_{\mu}= \bar{s} \gamma^{\mu}\left(1-\gamma_5 \right)c$ is the weak current for the $c\to s$ transition. The sums in the right-hand side run over all strange $K^*$ and charmed $D^*$ meson resonances that couple to the $K\pi$ and $D\pi$ two-particle systems, respectively\footnote{Charmed-strange $D_s^*$ resonances replacing the $D^*$ pole in the second term of Eq. (\ref{sumofcontributions}) are also possible. However, they are expected to be suppressed due that they lie farther away the physical region.}. In our model, we will neglect the contribution of the contact term $C$ (Figure \ref{FeynmanDiag}c). Furthermore, since we want to  test the hypothesis that the $D^*$-pole plays the role of the $S$-wave contribution, we will not consider the effects of the latter in our analysis. Other excited states like $K^*_0(1430)$ and $K_2^*(1430)$ also will be excluded in our analysis since they play a marginal role in the fit, according to Refs. \cite{BESIII1, BABAR}. 

 In the previous expression we have defined $D_{K^*}(s_{12}) = s_{12} - m_{K^*}^2 + im_{K^*}\Gamma_{K^*}(s_{12})$ and $D_{D*}(s') = s'-m_{D^*}^2$ (with $s'=(p-p_2)^2$), which arise from the $K^*$ and $D^*$ intermediate resonances, respectively.  The expressions for the $K^*$ energy-dependent decay width are provided in the following section. We do not consider the decay width in the $D^*$ propagator, because its off-shellness $s'<(m_D-m_\pi)^2$ is below the threshold for $D^*$ decays, thus its self-energy does not develop an imaginary part.

For the strong vertices  $\langle K \pi | K^{*}\rangle$ and $\langle D^{*} \pi | D\rangle $ we use the Feynman rule $ig_{V^*}\varepsilon_V\left ( P_V \right)\cdot  \left( q_1 -q_2 \right)$ following the flux momenta  convention $V^* \left(P_V, \varepsilon_V\right) \to P_1\left(q_1 \right)P_2\left( q_2\right)$
($V^*$ denotes a vector meson resonance and $P_i$ 
stand for the pseudoscalar mesons). Using this definition the $g_{V^*}$ couplings are real quantities.  However, in the sum of amplitudes we will assume a relative phase between the contributions  mediated by strange and charmed resonances, namely $g_{D^*}=|g_{D*}|e^{i\delta_{D^*}}$ as this is not defined between the two sectors unless we assume a flavor $SU(4)$ symmetry. 

  Finally,  we need the expressions for the hadronic matrix elements of the weak current entering the right-hand-side of Eq. (\ref{sumofcontributions}). Using the Lorentz covariance properties of these hadronic matrix elements we have (see \cite{CHANG2016, ABADA})

\vspace{0.05cm}
\begin{strip}
\begin{eqnarray}\label{CurrentK*}
\langle  K^*\left(\varepsilon ^*, P_{K^*}\right) |j_{\mu} | D\left(p\right) \rangle &=& \frac{2iV}{m_{D}+m_{K^*}}\epsilon_{\mu \nu \rho \sigma} \varepsilon ^{* \nu}p^{\rho} P_{K^*}^{\sigma} - 2 m_{K^*} A_0 \frac{q\cdot \varepsilon ^*}{q^2}q_{\mu} - \left( m_{D} + m_{K^*} \right) A_1\varepsilon ^{* \beta} T_{\beta \mu}\left( q \right) \nonumber \\
& \quad \quad +& 
A_2\frac{\varepsilon^* \cdot q}{m_D + m_{K^*}}\left(p+P_{K^*} \right)^{\beta}T_{\beta \mu}\left( q \right),
\end{eqnarray}

\begin{eqnarray}\label{CurrentD*}
\langle  K\left(p_1\right) | j_{\mu} | D^*\left(\varepsilon', p_{D^{*}}\right) \rangle &=& -\frac{2iV^{\prime}}{m_{1}+m_{D^{*}}}\epsilon_{\mu \nu \rho \sigma} \varepsilon'^{\nu}p_1^{\rho} P_{D^*}^{\sigma} - 2 m_{D^{*}} A_0^{\prime}\frac{q\cdot \varepsilon'}{q^2}q_{\mu} - \left( m_{1} + m_{D^{*}} \right) A_1^{\prime}\varepsilon'^{ \beta} T_{\beta \mu}\left( q \right) \nonumber \\
& \quad \quad -& 
A_2^{\prime}\frac{\varepsilon' \cdot q}{m_1 + m_{D^{*}}}\left(p_1+P_{D^{*}} \right)^{\beta}T_{\beta \mu}\left( q \right)\ ,  
\end{eqnarray}

\noindent where $\varepsilon^*(\varepsilon')$ denotes the polarization four-vector of the $K^* (D^*)$ vector meson and $T_{\alpha \beta}(q) \equiv g_{\alpha\beta} - q_{\alpha}q_{\beta}/q^2$ defines a transverse tensor such that $q^{\alpha}T_{\alpha \beta} = T_{\alpha \beta}q^{\beta}=0$. Note that we need four form factors $[V, A_0, A_1, A_2]$ (respectively $[V', A'_0, A'_1, A'_2]$) to describe the $D\to K^*$ ($D^* \to K)$ matrix elements. These two sets of form factors depend only upon the same variable, the square of the momentum transfer $q^2=L^2=(p_D-p_{K^*})^2=(p_{D^*}-p_K)^2$. Note that for the electronic modes $D_{e4}$ decays, the axial form factors $(A_0,\ A'_0)$ give negligible contributions.  


Using the expressions for the strong and hadronic weak vertices just defined, we can get the following expressions for the form factors of the $D_{\ell 4}$ decay defined in Eq. (\ref{most-general-decomposition})

\begin{eqnarray}\label{FFDef}
-\frac{i}{M}F 
&=&
-g_{K^*}\frac{{\rm BW}_{K^*}(s_{12})}{m_{K^*}^2} \left[(m_{K^*}+M)A_{1}\frac{P\cdot Q}{m_{K^*}^2} + \frac{2A_{2}Y}{m_{K^*}+M} \right] \nonumber \\ 
&\quad\quad +&  g_{D^*}\frac{{\rm BW}_{D^*}(s')}{m_{D^*}^2}\left[ (m_{D^{*}}+m_{1})A_{1}^{\prime}(Z+1)+\frac{XA_{2}^{\prime}}{m_{D^{*}}+m_{1}} \right]  
 -  2g_{S} \frac{{\rm BW}_S(s_{12})}{m_{S}^2}  f_{+}\left( q^2 \right), \nonumber\\
-\frac{i}{M}G 
&=& 
g_{K^*}\frac{{\rm BW}_{K^*}(s_{12})}{m_{K^*}^2}(m_{K^*}+M)A_{1} + g_{D^*}\frac{{\rm BW}_{D^*}(s')}{m_{D^*}^2}\left[ (m_{D^{*}}+m_{1})A_{1}^{\prime}(Z-1) + \frac{XA_{2}^{\prime}}{m_{D^{*}}+m_{1}} \right] \nonumber \\
-\frac{i}{M}R 
&=& 
g_{K^*}Y\frac{{\rm BW}_{K^*}(s_{12})}{m_{K^*}^2} \left[ \frac{2m_{K^*}A_{0}}{L^2} - (m_{K^*}+M)\frac{A_{1}}{L^2} + \frac{2A_{2}}{m_{K^*}+M}\frac{P\cdot L}{L^2} \right], \nonumber \\
&\quad \quad +& g_{D^*}\frac{{\rm BW}_{D^*}(s')}{m_{D^*}^2} \left[  \frac{2m_{D^{*}}A_{0}^{\prime}X}{L^2} + (m_{D^{*}}+m_{1})A_{1}^{\prime}\left( 2Z-\frac{X}{L^2} \right) - \frac{XA_{2}^{\prime}}{m_{D^{*}}+m_{1}} \frac{(P+Q)\cdot L}{L^2} \right] \nonumber \\
&\quad \quad -& g_{S} \frac{{\rm BW}_S(s_{12})}{m_{S}^2} \left[ f_+ \left( q^2 \right) + f_- \left( q^2 \right) \right] ,\nonumber \\
-\frac{H}{M^3} 
&=& 
2i\left[ g_{K^*}\frac{{\rm BW}_{K^*}(s_{12)}}{m_{K^*}^2} \frac{V}{M+m_{K^*}} - g_{D^*}\frac{{\rm BW}_{D^*}(s')}{m_{D^*}^2} \frac{V^{\prime}}{m_{D^{*}}+m_{1}} \right],
\end{eqnarray}
\end{strip}
\vspace{0.05cm}
\noindent where the quantities $X, \ Y$ and $Z$ are defined as 
\begin{eqnarray}
X&=& \left( p+P_{2}\right)\cdot L - \frac{1}{m_{D^{*}}^{2}}\left[\left( p+P_{2}\right)\cdot p_{D^{*}}\right]  L \cdot p_{D^*},  \nonumber \\
Y&=& Q\cdot L -\frac{1}{m_{K^*}^{2}}\left( P\cdot Q\right)\left(P\cdot L\right),  \nonumber \\
Z&=&\frac{1}{2}\left[1- \frac{1}{m_{D^*}^{2}}  \left( p+P_{2}\right)\cdot p_{D^*} \right] + 1. 
\end{eqnarray}
 In Eq. (\ref{FFDef}) we have introduced the Breit-Wigner shapes BW$_{K^*}(s_{12})=m_{K^*}^2/D_{K^*}(s_{12})$, BW$_{D^*}(s')=m_{D^*}^2/D_{D^*}(s')$ for vector mesons and BW$_S(s_{12})$ for the scalar strange meson. Each of the form factors in Eqs. (\ref{FFDef}) is written as the sum of the vector ($K^*\ ,D^*)$ and scalar meson pole contributions. Although we do not consider the effects of the scalar $K_0^*(700)$ in this analysis, we included its contributions in the expressions for the form factors $F$ and $R$: 
\begin{eqnarray}
\langle  K_0^*\left( p_{K^*}\right) |j_{\mu} | D\left(p\right) \rangle &=& f_+(q^2)(p+p_{K^*})_{\mu} \nonumber \\ && \ \ \ + f_{-}(q^2)(p-p_{K^*})_{\mu}\ .
\end{eqnarray}

Following the analyses  carried out by the BESIII experiment \cite{BESIII1, BESIIID0}, we use a monopolar function for the $q^2$-dependence of the weak form factors, namely ($i=0,1,2$)

\begin{eqnarray} \label{SPD}
A_{i}(q^2) &=& \frac{A_{i}\left( 0 \right)}{1-q^2/m^2_{Ai}},\nonumber\\
V\left(q^2\right) &=& \frac{V\left( 0 \right)}{1-q^2/m_{V}^2}, \nonumber\\
A^{\prime}_{i}(q^2) &=& \frac{A^{\prime}_{i}\left( 0 \right)}{1-q^2/m^{ 2}_{Ai}}  ,\nonumber\\
V^{\prime}\left(q^2\right) &=& \frac{V^{\prime}\left(0 \right)}{1-q^2/m^{2}_{V}}\ .
\end{eqnarray}
\noindent  These form factors depend on two parameters: their normalization at $q^2=0$ and on the values of pole masses. For the latter, it is common to use the masses of the lighter resonances: $m_V$ ($m_{A_i}$) for the vector (axial vector) resonance, with the appropriate quantum numbers to describe the $D\to K^*$ and $D^*\to K$ vertices. 

 The values required for the pole masses and the form factors at $q^2=0$ are shown in Table \ref{TableFFValues}.
 In the case of the $D^* \to K$ transition, the form factors at $q^2=0$ were calculated in Ref. \cite{SGK} using  the model described in Ref. \cite{WirbelStechBauer,  WirbelBauer}.   
 This model \cite{WirbelStechBauer,  WirbelBauer} also assumes a monopole-type behavior of the form factors for the momentum transfer dependence, namely, with nearest-pole dominance.  The pole masses correspond to the nearest resonances that couple to the weak currents with appropriate quantum numbers.  Using this model for the subleading $D^*\to K$ transition, we are left with the form factors for the dominant $D\to K^*$ contribution at $q^2 =0$ as the only undetermined constants to be fitted by data.

\begin{table}[hbt!] 
\centering
\begin{tabular}{|c|c|c|c|}
\hline
\multirow{3}{6em}{\centering Transition} & \multirow{3}{4em}{\centering Form factor} & \multirow{3}{6em}{\centering Form factor at $q^2=0$} & \multirow{3}{6em}{\centering Pole mass [GeV]} \\ 
 & & & \\
  & & & \\
 \hline
\multirow{4}{6em}{\centering  $D \to K^*$ \cite{WirbelStechBauer, WirbelBauer}}
  & $A_{1}$ & 0.619(17) & 2.60 \\
  & $A_{2}$ & 0.487(30) & 2.60 \\
  & V & 0.873(43) & 1.80 \\
 \hline
 \multirow{4}{4em}{\centering $D\to K^*$ \cite{PhysRevD.101.013004}} & $A_{0}$ & 0.629 & 1.968 \\
 & $A_{1}$ & 0.596 & 2.010 \\
 & $A_{2}$ & 0.540 & 2.010 \\
 & $V$ & 0.959 & 2.112  \\
\hline
\multirow{4}{6em}{\centering $D^* \to K$ \cite{SGK} } & $A_{0}^{\prime}$ & 0.78 & 1.97 \\
 & $A_{1}^{\prime}$ & 1.02 & 2.53 \\
 &$ A_{2}^{\prime}$ & 0.40 & 2.53 \\
 & $V^{\prime}$ & 0.90 & 2.11 \\
\hline
\end{tabular}
\caption{\\ Weak form factors at zero momentum transfer and pole mases  for $D\to K^*$ and $D^*\to K$ transitions. In the limit of isospin symmetry, they are the same for both charge states of each $D \to K^*$ and $D^*\to K$ transition. References \cite{SGK, PhysRevD.101.013004} do not quote uncertainties for their calculated values, therefore we used only \cite{WirbelStechBauer, WirbelBauer} to estimate our uncertainty on $|V_{cs}|$. 
}\label{TableFFValues}
\end{table}

In summary, for $D_{e4}$ decays, the observables would depend upon 6 normalization factors and 6 pole masses that arise from the weak  hadronic matrix elements. Since the $D^*$ pole is a subleading contribution, we will use the predicted values of the weak form factors $V', A'_1, A'_2$ as well as the meaured  pole masses as input parameters in the fit. 
In addition, we have two strong ($DD^*\pi$ and $KK^*\pi$) coupling constants and one relative phase among them. Since the $g_{K^*}$ strong coupling is fixed from the decay width of $K^*(892)$ meson, we will leave as a free parameter only $g_{D^*}$ and its relative phase. Therefore, we are left with only five parameters ($V(0), A_1(0), A_2(0), |g_{D^*}|$ and $\delta_{D^*}$) to be fitted from data in addition to the mass and width of the $K^*(892)$. Alternatively, we can factorize $|A_1(0)|^2$ as a normalization of the decay observables, and use instead $r_V\equiv V(0)/A_1(0)$ and $r_2\equiv A_2(0)/A_1(0)$ as the free parameters for the hadronic weak vertex. Other mass parameters needed are taken from Ref. \cite{PDG}.

\subsection{Resonant shapes}\label{sb}

The $K\pi$ system in $D_{\ell 4}$ decays can be produced in a resonant way because the accessible kinematical region $m_K+m_{\pi} \leq \sqrt{s_{12} } \leq m_D -m_{\ell}$ is populated by the vector and scalar resonances. Since these resonances provide the dominant contribution, it is convenient to fix our convention for the Breit Wigner functions. In this paper we keep only the lowest lying resonance and use the following parametrization
\begin{equation}
{\rm BW}_{K^*}\left(s_{12}\right) = \frac{m_{K^{*}}^2}{s_{12}-m_{K^{*}}^2+im_{K^{*}} \Gamma_{K^*} \left(s_{12}\right)} \ .
 \end{equation}
The parametrization used by the BESIII colaboration in Refs \cite{BESIII1, BESIIID0} has a similar form. However, while our decay width, which is calculated using Feynman rule for the $VPP$ vertex defined at the beginning of Section  \ref{FF}, is given by:
\begin{equation}
\Gamma_{K^*} \left( s_{12} \right) =\Gamma_{K^*}\frac{m_{K^*}^4}{s^{2}_{12}}\frac{\lambda^{3/2}_{12}}{\lambda^{3/2}_{K^*}}\Theta\left[ s_{12}-\left( m_K+m_{\pi} \right)^2 \right] \ , 
\end{equation}
the expression used in Refs. \cite{BESIII1, BESIIID0} reads
\begin{equation}\label{wBES}
\Gamma_{{\rm BESIII}}(s_{12})=\Gamma (s_{12})\cdot \left( \frac{B_1(p)}{B_1(p^*)}\right)^2 \ . 
\end{equation} 

\begin{figure}
\centering
\includegraphics[width=8.5cm]{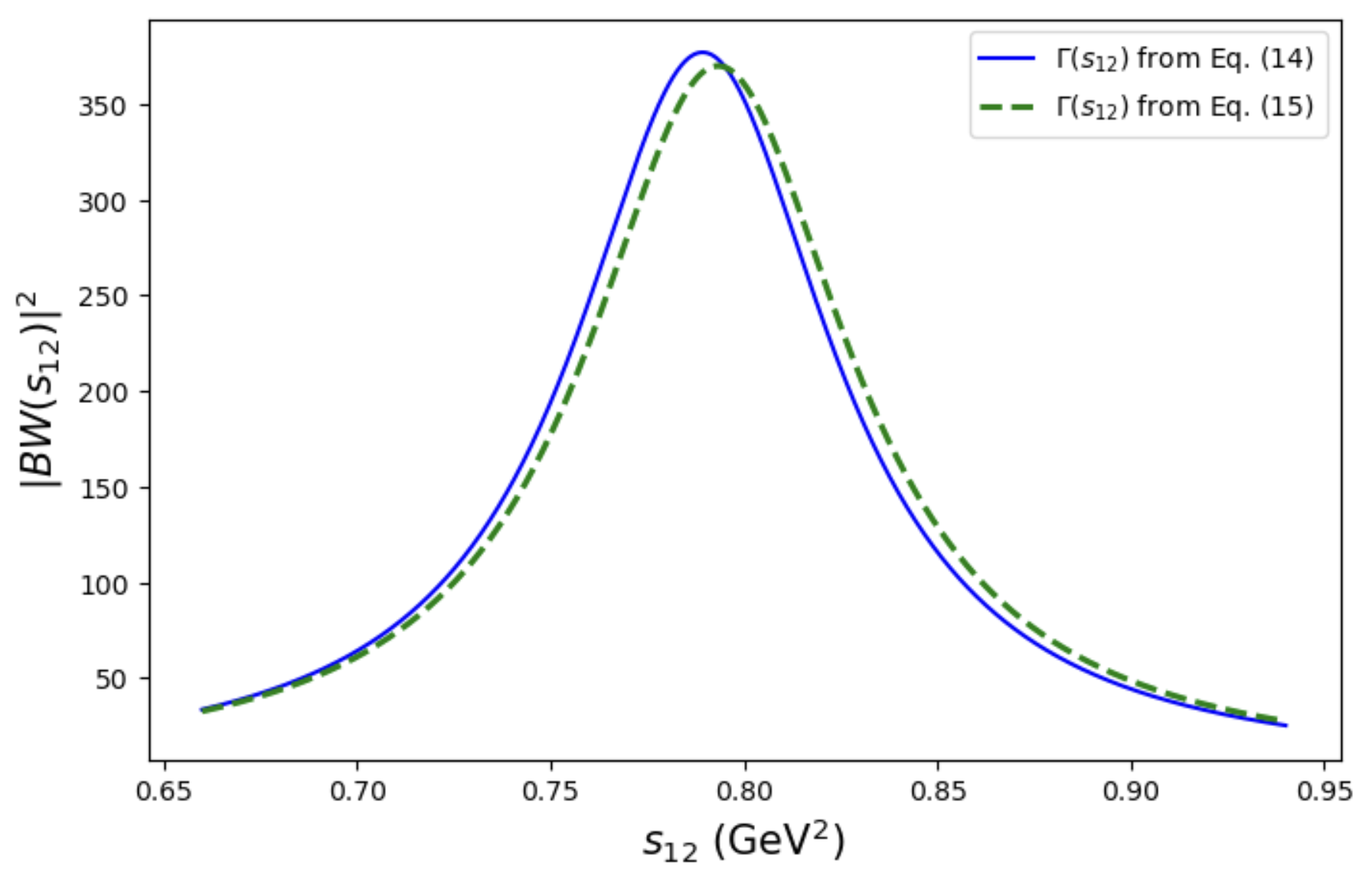} 
  \caption{ Comparison of the Breit-Wigner shapes of the dominant $K^*(892)$ resonance used in this work (solid-line)  and the one used in Ref. \cite{BESIII1} (dashed-line). }
  \label{BWF}
\end{figure}

\begin{table}[H] 
\centering
\begin{tabular}{|c||c|c|}
\hline
& & \\
$D^+_{\ell 4}$ Channel & $\Gamma (K^{*0})$  & $\Gamma (K^{*0}+D^{*0})$ \\ 
 & ($\times 10^{-11}$ MeV ) & ($\times 10^{-11}$ MeV)  \\
& &  \\
 \hline
 \hline 
&  & \\  
$\ell = e$ \cite{WirbelStechBauer, WirbelBauer}&  2.26(13)  & 2.56(15) \\ 
& \big[2.15(13)\big] & \big[2.26(13)\big]\\
&&\\
\hline
&  & \\
$\ell = \mu $ \cite{BESIII1} &  2.09(12)  & 2.37(13) \\ 
& \big[1.98(12)\big] & \big[2.06(12)\big]\\
&  &\\
\hline
&  & \\
$\ell = e $ \cite{PhysRevD.101.013004} &  2.09(10)  & 2.42(19) \\ 
& \big[1.94(10)\big] & \big[2.05(10)\big]\\
&  &\\
\hline
&  & \\
$\ell = \mu $ \cite{PhysRevD.101.013004} &  1.98(10)  & 2.30(17) \\ 
& \big[1.84(9)\big] & \big[1.95(10)\big]\\
&  &\\
\hline\hline
\end{tabular}
\caption{\\ Decay rates of $D^+ \to K^-\pi^+\ell^+\nu_{\ell}$ channels. Results within square brackets indicate that the $m_{K \pi}$ invariant mass was integrated over the $K^*(892)$ dominated region, $\left[ 0.8,1.0 \right]$ GeV. The second (third) column contains the effect of the $K^*(892)$ ($K^*(892)+D^*$) resonance(s). The $g_{D^*}$ constant is assumed to be a real parameter ($|g_{D^*}|=8.39(8)$ extracted from the measured partial width \cite{PDG}).
} \label{DRate}
\end{table}

The previous expression contains an additional term which depends on $B_1(p) = \left[ 1+r_{BW}^2p^2 \right]^{-1}$, the so-called Blatt-Weiskopf damping factor. In the above expressions we have defined $\lambda\left(  a,b,c,\right) = a^2+b^2+c^2-2\left( ab+ac+bc \right)$ and $p= \lambda^{1/2}\left(s_{12},m_{K}^2, m_{\pi}^2 \right)/(2\sqrt{s_{12}}),\ p^*=p(\sqrt{s_{12}}=m_{K^*})$. The last factor in the right-hand-side of Eq. (\ref{wBES}) becomes unity at the resonance peak, $\lambda_{12}=\lambda(s_{12},m_K^2,m_{\pi}^2)$. It distorts slightly the Breit-Wigner shape in the kinematical range around the $K^*(892)$ resonance, as it can be observed in Figure \ref{BWF}. 

In order to assess the difference expected between the muonic and electronic $D_{\ell4}$ decay channels as well as the effects of the $D^*$ pole resonance, we have evaluated the rates for $D^+ \to K^-\pi^+\ell^+\nu_{\ell}$ decays (see Table \ref{DRate}).  We provide results obtained using the two different sets of form factors at zero momentum transfer indicated in Table \ref{TableFFValues}. We use as input values  $|V_{cs}|= 0.975 \pm 0.006$ \cite{PDG}, as well as the meson masses quoted in the PDG \cite{PDG}. It can be observed that  considering the $D^*$ pole increases de decay rate by about 12\% (16\%) using model \cite{BESIII1} (model of Ref. \cite{PhysRevD.101.013004}). On the other hand, the model of Ref. \cite{PhysRevD.101.013004} gives lower rates for the decays under consideration either, using one ($K^*$) or two ($K^*+D^*$) resonances.  The rates for $D^+_{\mu 4}$ decays are smaller than for $D^+_{e4}$, as expected from phase-space considerations. Those results are in good agreement with current world averages given (as branching fractions) in Eq. (\ref{ExpBR})

\section{Fit to BESIII data}\label{FitCriteria}
Once we have estimated the effects of different weak form factor models and of the strange and charmed pole contributions on the rates of $D^+_{\ell 4}$ decays, we proceed to fit the data reported by the BESIII colaboration \cite{BESIII1}. We consider the datasets available for charged or neutral $D$ mesons decays: the $K\pi$ mass distribution ($d\Gamma _{} / d m_{K \pi}$), the momentum   transfer distribution ($d\Gamma _{\rm exp} / d q^2$), the branching fractions (${\cal BR}_{\rm exp}$) and the helicity form factors distributions ($|H_{\pm, 0 \  {\rm exp}}(q^2)|^2$)  in the case of $D_{e4}^+$ decays.  
The helicity form factors are defined in terms of the weak form factors  as follows \cite{Korner:1989qb} (see also \cite{FOCUS:2002lsy}):
\vspace{0.01cm}
\begin{strip}
\begin{align}
H_{\pm}(q^2) &= \left( m_D + \sqrt{s_{12}} \right)A_1 \left(q^2\right) \mp \frac{2m_D p_{K\pi}}{m_D + \sqrt{s_{12}} }V\left(q^2\right)\nonumber \\
H_0(q^2) &=  \frac{1}{2q\sqrt{s_{12}}}\left[ \left(m_D^2-s_{12}-q^2 \right)\left( m_D + \sqrt{s_{12}} \right) A_1\left( q^2 \right) - 4\frac{m_D^2 p_{K\pi}^2}{m_D + \sqrt{s_{12}}}A_2\left( q^2 \right)\right],
\end{align}

\vspace{0.01cm}
where $p_{K\pi} =\lambda ^{1/2}(m_D^2,s_{12},q^2)/2m_D$ is the momentum of the $K\pi$ system in the $D$ meson rest frame. 
 The $\chi ^2$ function is built  as follows

\begin{eqnarray}\label{chi_square}
\chi^2 &=&  \sum_{i} \left( \frac{\frac{d\Gamma ^{i}_{th}} {d \sqrt{s_{12}}}  - \frac{d\Gamma ^{i}_{exp}}{d m_{K \pi}}}{ \sigma ^{i}_{m_{K\pi},exp}}\right)^2 + \sum_{j} \left(\frac{\frac{d\Gamma ^{j}_{th}} {d s_{34}}  - \frac{d\Gamma ^{i}_{exp}}{d q^2}}{ \sigma ^{j}_{q^2,exp}} \right)^2 +\left( \frac{\mathcal{BR}_{th} - \mathcal{BR}_{exp}}{ \sigma _{\mathcal{BR}, exp}}\right)^2 \nonumber \\
& \quad \quad +&  \sum_{k} \left(  \frac{\lvert H_{+,th}^{k}\left( q^2 \right)\rvert ^{2} - \lvert H_{+,exp}^{k}\left( q^2 \right)\rvert ^{2}}{\sigma^{k,exp}_{H_{+}^{2}}}\right)^2 + \sum_{l} \left(  \frac{q^2\lvert H_{0,th}\left( q^2 \right)\rvert ^{2} - q^2\lvert H_{0,exp}\left( q^2 \right)\rvert ^{2}}{\sigma^{l,exp}_{q^2\lvert H_{0}\rvert ^{2}}}\right)^2 .
\end{eqnarray}
\end{strip}
The denominator in each term denotes the experimental error for the associated observable and the sum is carried over each set of bins reported by the experiment.

 Since data on $D_{e4}$ decays provided by the BESIII collaboration \cite{BESIII1, BESIIID0} are not de-convoluted from detector effects, we have fitted our model to the central points reproduced from their best fitted curves,  but we keep the experimental error associated to the bins of each observable. Certainly,  this is a limitation in our procedure given the impossibility to access the information on folding effects. On the other hand, the seven free parameters of our fit are the following: $(m_{K^*},\ \Gamma_{K^*},\ |A_{1}(0)|, \ r_2., \ r_V)$ and the phase and magnitude of the effective $g_{D^*}$ coupling. The parameters of subleading $D^*$-pole  contributions (weak form factors) are taken from theoretical predictions  and/or the PDG compilation  \cite{PDG}. The experimental uncertainties were treated as uncorrelated, since we had not access to the covariance matrix.

\section{Fit results}\label{RyD}

In this section we present the results of our fits to experimental data on $D^+ \to K^-\pi^+ e^+ \nu_{e}$ \cite{BESIII1} and $D^0 \to \overline{K}^0\pi^- e^+ \nu_{e}$ \cite{BESIIID0} decays ($D^+_{e4}$ and $D^0_{e4}$ for short, respectively). The hadronic matrix elements of these two decays are related by isospin symmetry, namely: $\langle K^- \pi^+| J_{\mu}| D^+ \rangle =- \langle \overline{K^0}\pi^- | J_{\mu}| D^0 \rangle$. We also provide the branching fractions of these decays in the full and the $K^*(892)$ kinematical regions of the $K\pi$ invariant mass. The results of the fits are shown in Tables \ref{RF1} and \ref{RF1D0}, respectively, for $D^+_{e4}$ and $D^0_{e4}$ decays. 

 Table \ref{RF1} displays the results for the  $D^{+}\to K^{-}\pi^{+}e^{+}\nu_{e}$ channel. The second column shows the values quoted  by the BESIII collaboration \cite{BESIII1},  using a model where the $S$-wave background and the vector $K^*(892)$ are included in the fit.  
Column three corresponds to our results. In our fits we have fixed the CKM matrix element to the value quoted in Ref. \cite{PDG}, namely $|V_{cs}|=0.975\pm 0.006$.  

We observe that the presence of the $D^*$ pole is mainly correlated with values of the form factors at zero momentum transfer, while the $K^*(892)$ resonance parameters remain basically the same.  Also,  the extracted value for the magnitude of the $D D^* \pi$ coupling,  is in good agreement with the value obtained from the experimental width of $D^* \to D\pi$ \cite{PDG} decays (indicated within square brackets in the third column of Table \ref{RF1}).  Finally,  the values of the ratios of form factors $r_V, r_2$ are in good agreement with the ones obtained in Ref.  \cite{BESIII1} 

Table \ref{RF1D0} presents our results for the  $D^{0}\to \bar{K}^{0}\pi^{-}e^{+}\nu_{e}$ channel using the data reported in reference \cite{BESIIID0}. The analysis of the BESIII collaboration \cite{BESIIID0} in this case is similar to the $D^+ \to K^-\pi^+ e^+\nu_{e}$ decay (namely, the dominance of the vector $K^{*-}(892)$ resonance and the $S$-wave background in the $K\pi$ system is assumed). The observed number of events is approximately a factor of two lower than for the charged $D^+_{e4}$ decays. In the second column we display the results of the fit obtained by the BESIII collaboration \cite{BESIIID0}. In the third column we show the results of our fit. Similarly to the $D_{e4}^+$ case, the largest variations with respect to the fit of Ref. \cite{BESIIID0} are observed in the weak form factors at zero momentum transfer. The magnitude of the $D^*D\pi$ coupling in this case is a bit larger (but compatible) compared to the one of its charged counterpart, as expected from isospin symmetry, but the relative phase between the $D^*$ and $K^*$ contributions is different.

\begin{table}[hbt!] 
\centering
\begin{tabular}{ c c c }
\hline\hline
& & \\
\multirow{2}{5em}{\centering Parameter} & BESIII fit & This work \\ 
& ($K^{*0}$+ S-Wave) & ($K^{*0}$+$D^{*0}$) \\ 
& & \\ 
 \hline
& & \\ 
$m_{K^{*0}\left(892 \right)}$ & \multirow{2}{5.8em}{\centering 894.60(26)} & \multirow{2}{5.8em}{\centering 892.95(1.23)} \\ 
\scriptsize $[$MeV$]$ & & \\ 
& & \\ 
$\Gamma_{K^{*0}\left(892 \right)} $ & \multirow{2}{5.8em}{\centering 46.42(56)} & \multirow{2}{5.8em}{\centering 45.38(1.20)} \\ 
\scriptsize $[$MeV$]$ & & \\ 
& & \\ 
$A_{1}\left(0\right)$ &  0.619(17) & 0.627(10) \\ 
& & \\ 
$r_2 $ & 0.788(43) & 0.766(55) \\ 
& & \\ 
$r_V $  &  1.411(58) & 1.453(77) \\ 
& & \\ 
$\delta _{g_{D^{*}}}$ [rad] & - & 0.643(156) \\ 
& & \\ 
$| g_{D^{*}} |$ & - & 8.744(317) \\ 
& & \big[ 8.390(80) \big] \\ 
& & \\ 
$\chi^2/$n.d.f. & 1.006 & 0.78 \\ 
& & \\ 
\hline\hline
\end{tabular}
\caption{\\  Fits to  $D^+ \to K^-\pi^+ e^+\nu_e$  data from the BESIII collaboration \cite{BESIII1}.  The second column shows the fit results of Ref. \cite{BESIII1}. Third to fifth columns are the results of our fits. The last two columns use two reference input values for the axial form factor ($A_1(0)=0.619$ (BESIII) and $A_1(0)=0.629$).  Our results when $A_1(0)$ is also allowed to float are shown in third column. The magnitude of the $g_{D^{*}}$ parameter extracted from the  experimental $D^*$ partial widths is given within squared brackets.
} \label{RF1}
\end{table} 

Once all the free parameters have been fixed from experimenal data as done in columns 3 of Tables \ref{RF1} and \ref{RF1D0}, we can predict the branching fractions of $D^{+,0}_{e4}$ decays. In Table \ref{BFr} we show in the second column the branching fraction calculated in our model when we integrate the $K^*+D^*$ pole  contributions over the full kinematical region. In the following two columns we quote the results obtained when we restrict the $K\pi$ invariant mass to lie in the $K^*(892)$ kinematical region, namely $m_{K^*} -\Gamma_{K^*} \leq \sqrt{s_{12}} \leq m_{K^*} +\Gamma_{K^*}$, and we allow for the  two pole $K^*+D^*$ (third column) and the single $K^*$ resonance (fourth column) contributions, respectively. Finally, in the fifth column we display our results when only the $D^*$ is allowed to contribute in the full kinematical region. 

\begin{table*}[hbt!] 
\centering
\begin{tabular}{| c| c| c c| c |}
\hline\hline
& & & & \\
 & Total BR  & \multicolumn{2}{c}{$K^*$ region }\vline & Full region  \\ 
{\centering Channel} & & & &\\
& ($K^{*}+D^*)$ & $(K^*+D^*)$ & ($K^*$ only) &  ($D^*$ only)  \\
& & & & \\
 \hline
& & & &  \\
$D^+\to K^-\pi^+ e^+\nu_e$ & $4.12(24)\times 10^{-2}$  & $3.51(21)\times 10^{-2}\ \ $ \vline & $3.53(22)\times 10^{-2}$ & $4.05(36)\times 10^{-3}$  \\ 
& & & &  \\
& & & &$3.12(28)\times 10^{-3}$ ($D^*$+I)  \\
& & & &  \\
 $D^0\to \overline{K^0}\pi^- e^+\nu_e$& $1.565(86)\times 10^{-2}$ & $1.394(78) \times 10^{-2}\ $ \vline & $1.345(77)\times 10^{-2}$ & $1.71(16)\times 10^{-3}$  \\
&  & & &  \\
\hline\hline
\end{tabular}
\caption{\\  Branching fractions of $D_{e4}$ decays calculated in this work (second column). The third (fourth) column shows predictions when the rates are calculated in the $K^*(892)$ kinematical region $m_{K^*}-2\Gamma_{K^*} \leq \sqrt{s_{12}} \leq m_{K^*}+2\Gamma_{K^*} $ in  the two resonance (respectively, only $K^*$) model. The fifth column refers to the branching fraction stemming only from the $D^*$ pole contribution, except for the $D^*+I$ value which also includes the interference with the $K^*$ resonance. } \label{BFr}
\end{table*}

\begin{figure}[hbt!]
\centering
\includegraphics[width=9.5cm]{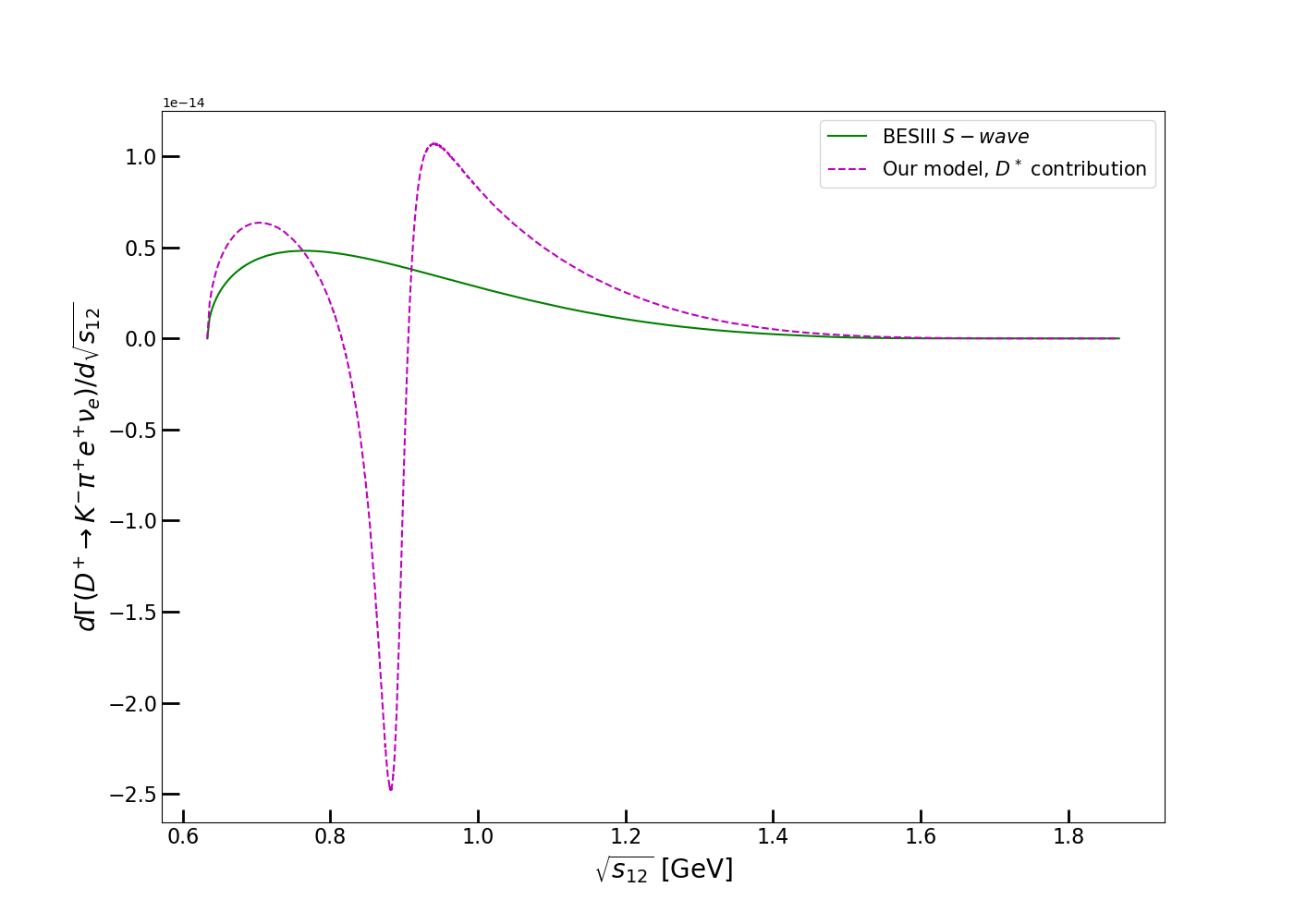}
  \caption{Invariant mass distribution of the $K^{-}\pi^{+}$ system in $D^+_{e4}$ decays: non-resonant $S$-wave contribution (green/solid line) \cite{BESIII1} and $D^*$ contribution + interference with $K^*$ (magenta/dashed line).} 
  \label{B3pTM}
  \end{figure}

From Table \ref{BFr} we can observe that the branching fractions calculated in our model with ($K^*+D^*$) poles, are in very good agreement with the world average of measured values shown in Eq. (\ref{ExpBR}). The contribution of the $D^*$ pole (including its interference with the $K^*(892)$ resonance), is approximately 7.6 \% (denoted as $D^*+I$ in the last column of Table \ref{BFr} of the total branching fraction. This $D^*$-pole contribution to the decay rate appears to be a bit larger compared to the $S$-wave contribution ($\sim 6$ \%) obtained by the BESIII collaboration \cite{BESIII1, BESIIID0}. We attribute this difference to the fact that the $D^*$-pole can in principle contribute to different wave configurations and can interfere with the dominant $P$-wave of the $K\pi$ system.

\begin{table}[hbt!] 
\centering
\begin{tabular}{c c c}
\hline\hline 
\multirow{4}{7em}{\centering Parameter} & \multirow{4}{7em}{\centering BESIII value (K$^{*-}$+S-Wave)} & \multirow{4}{7em}{\centering This work (K$^{*-}$+D$^{*+}$)} \\
& & \\
& & \\
& & \\
\hline\hline 
& & \\  
$m_{K^{*-}\left(892 \right)}$ & \multirow{2}{7em}{\centering 892.30(54)} & \multirow{2}{9em}{\centering 890.05(1.59)} \\ 
\scriptsize $[$MeV$]$ & & \\
& & \\
$\Gamma_{K^{*-}\left(892 \right)} $ & \multirow{2}{7em}{\centering 46.50(82)} & \multirow{2}{9em}{\centering 48.12(1.70)}\\ 
\scriptsize $[$MeV$]$ & &\\
& & \\
$A_{1}\left(0\right)$ & 0.610(8) & 0.590(12) \\ 
& & \\
$r_2 $ & 0.700(45) & 0.680(70) \\ 
& & \\
$r_V $  & 1.480(54) & 1.459(100) \\ 
& & \\
$\delta _{g_{D^{*}}} $ & - & 3.846(250) \\
& & \\
$| g_{D^{*}} |$ & - & 9.413(769) \\
& & \big[ 8.390(80) \big] \\
& & \\
$\chi^2/$n.d.f. & 1.126 & 0.746\\
& & \\
\hline\hline
\end{tabular}
\caption{Fit to $D^0 \to \overline{K^0}\pi^- e^+\nu_e$ data from BESIII experiment \cite{BESIIID0}. Results of the BESIII fit are shown in  the second column. Third column displays the results of our fit. The magnitude of  $g_{D^{*}}$ extracted from the experimental width is shown within squared brackets.}\label{RF1D0}
\end{table}

\begin{figure}[hbt!]
\centering
\includegraphics[width=9cm]{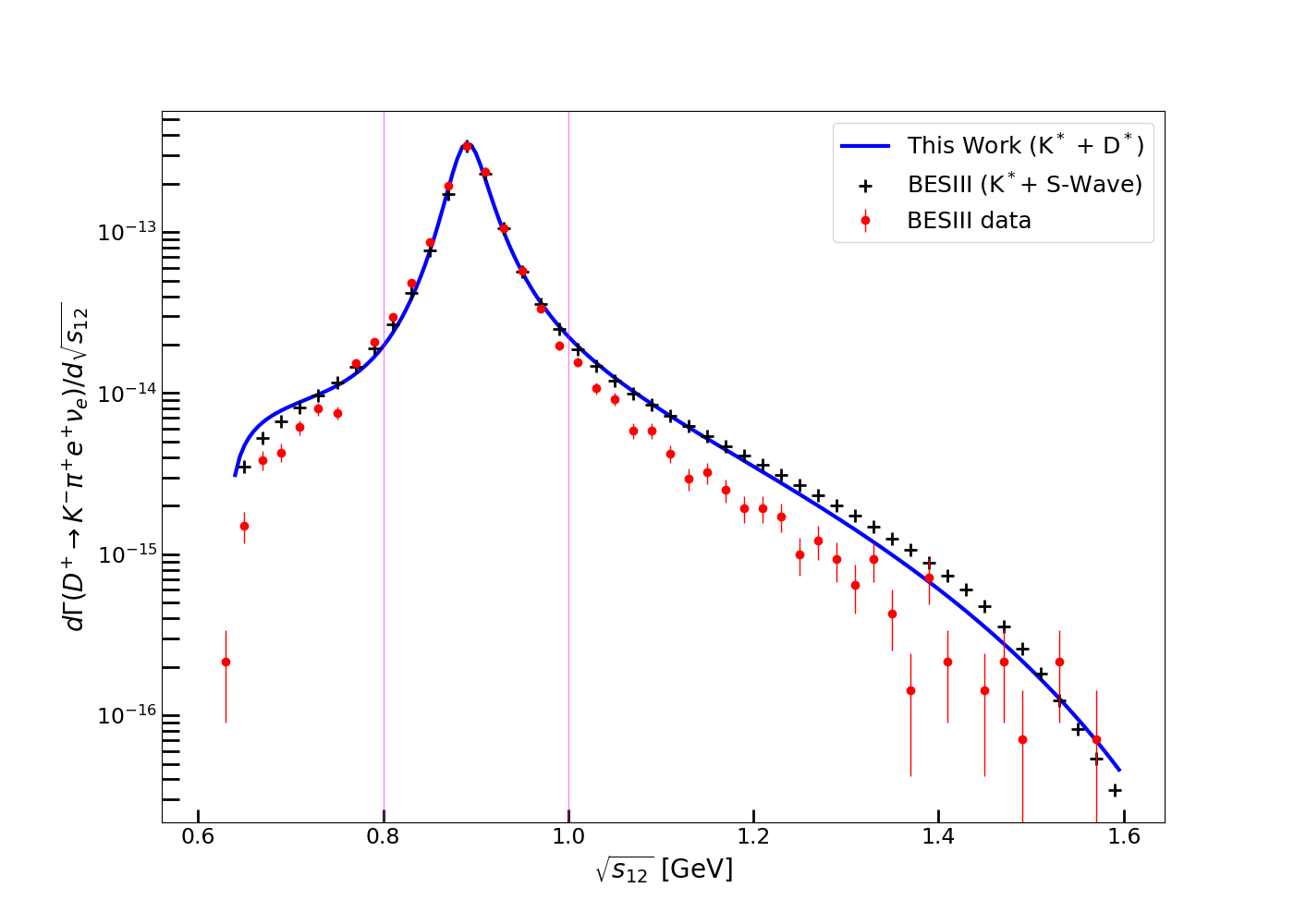}
  \caption{Invariant mass distribution for the $K^{-}\pi^{+}$ system in $D^+_{e4}$ decays. The blue-solid line corresponds to the best fit with our model and the black-crossed line corresponds to the BESIII model \cite{BESIII1}.  Folded experimental data from Ref. \cite{BESIII1} are shown with error bars.} 
  \label{KPip}
  \end{figure}

In Figures \ref{KPip} and \ref{RFmD0} 
 we plot, respectively, the invariant mass distributions of the $K\pi$  system in $D^+_{e4}$ and $D^0_{e4}$ decays. In Figure \ref{KPip} we compare the result of our fit (solid line) with the one (dotted line) obtained by the BESIII collaboration  \cite{BESIII1}.   The difference of the fits with experimental data, visible below and above the $K^*(892)$ resonance region, is because data has not been unfolded from detector effects.   Figure \ref{RFmD0} compares the results of our fit for the $K^0\pi^-$ mass distribution with the corresponding data on $D^0_{e4}$ decays.

\begin{figure}[hbt!]
\centering
\includegraphics[width=9cm]{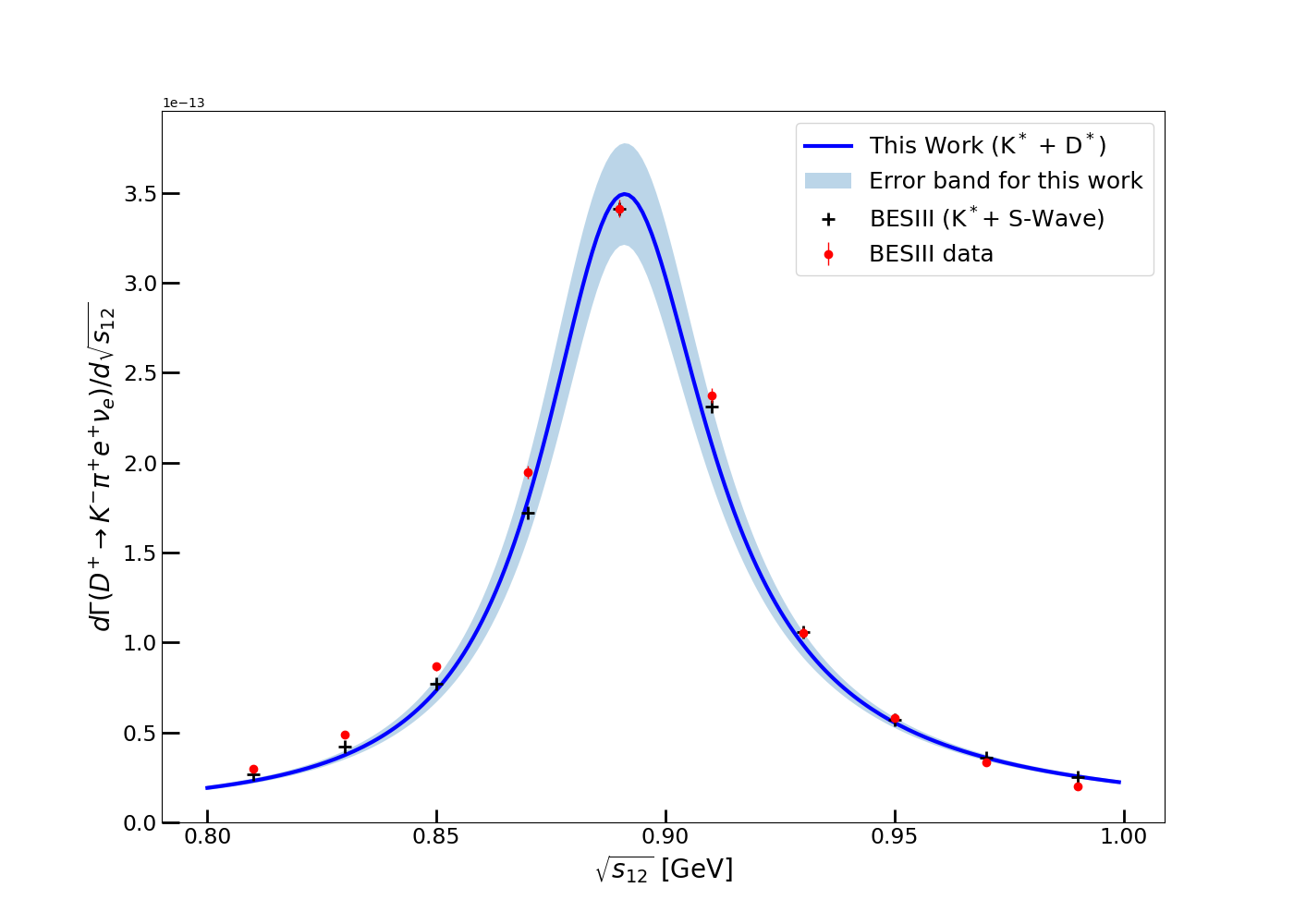}
  \caption{Invariant mass distribution of the $K^{-}\pi^{+}$ system in $D^+_{e4}$ decays. The blue-solid line represents the best fit from our model, while the shaded band indicates its uncertainty at the $95\%$ confidence level. The cross black points correspond to the BESIII model \cite{BESIII1}.  Unfolded experimental data from Ref. \cite{BESIII1} are shown with error bars. }
  \label{KPip_dr}
  \end{figure}

\begin{figure}[hbt!]
\centering
\includegraphics[width=9cm]{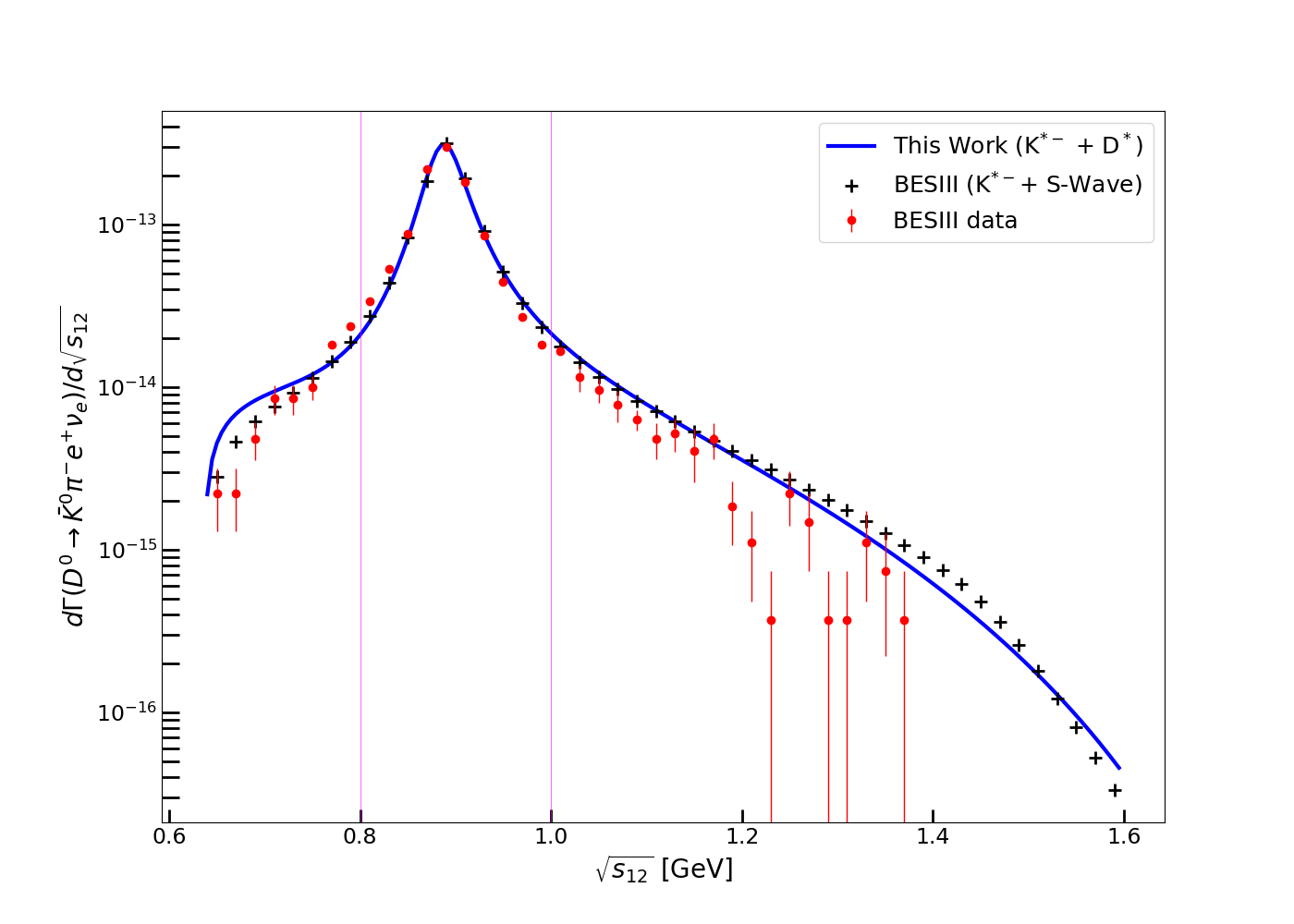}
  \caption{Invariant mass distribution for the $\bar{K}^{0}\pi^{-}$ system in $D^0_{e4}$ decays. The blue-solid line corresponds to the best fit within our model,  compared to data and best fit (line with crosses) from the BESIII collaboration \cite{BESIIID0} The vertical lines in magenta show the $K^{*-}$ dominated region.}
  \label{RFmD0}
\end{figure}

\begin{figure}[hbt!]
\centering
\includegraphics[width=9cm]{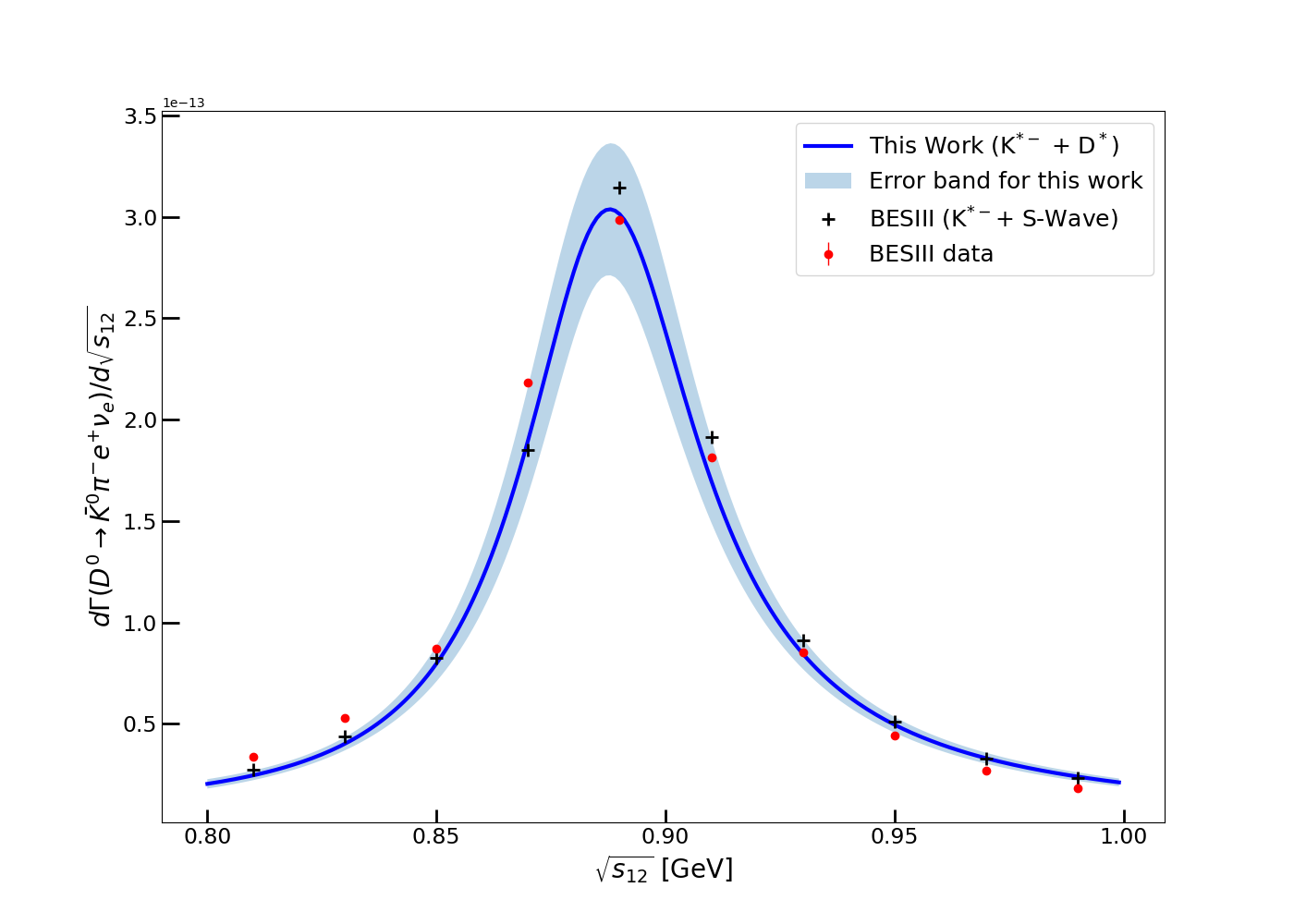}
  \caption{Invariant mass distribution of the $\bar{K}^{0}\pi^{-}$ system in $D^0_{e4}$ decays. The blue-solid line represents the best fit within our model while the shaded band indicates its uncertainty at the $95\%$ confidence level, compared to data and best fit from the BESIII collaboration \cite{BESIIID0}.}
\end{figure}

A comparison of the results of our fits with the ones provided in Refs. \cite{BESIII1, BESIIID0}, shows that the role of the $S$-wave configuration for the $K\pi$ system can be mimicked in our model by the $D^*$ pole contribution. The presence of the $D^*$ pole contributions does not affect the determination of the $K^*$ resonance parameters extracted from $D_{e4}$ decays. However, the $D^*$ pole contribution, affects mainly the determination of the weak charges of the $D\to K^*$ transition. The $D^*$ pole affects the determination of the branching fraction at the level of 7.6\%, which is similar to the contribution attributed to the $S$-wave contribution in Ref. \cite{BESIII1}.

While the $S$-wave configuration of the $K\pi$ system does not interfere with the $K^*$ contribution, this is not the case for the $D^*$-pole amplitude. This is clearly manifest in the $K^-\pi^+$ invariant mass distribution of $D_{e4}^+$ decays, as shown in Figure \ref{B3pTM}. The interference of the $K^*$ and $D^*$ contributions is responsible for the small reduction in the mass and width of the $K^*(892)$ (third column in Table \ref{RF1}) resonance as compared with the values obtained in Ref. \cite{BESIII1} (second colummn), without affecting the weak couplings of the $D\to K^*$ transition. Despite the different behavior of the $D^*$-pole and non-resonant $S$-wave contributions, their contribution to the branching fraction are very similar.

A final remark is in order. As pointed out in previous experiments \cite{BESIII1, BABAR}, scalar strange resonances ($\kappa(700), K_0^*(1430)$) give a negligible contribution to the $S$-wave amplitude. We have attempted a fit to the $D_{e4}^+$ data by adding simultaneously the $D^*$-pole and $\kappa(700) (\to K^-\pi^+)$ scalar contributions to the dominant $K^*(892)$ resonance. By fixing the mass and width of the $\kappa(700)$ to the average values in \cite{PDG}, we can achieve a good fit, to the expense of increasing the mass and width of the $K^*(892)$ to unaceptably large values ($m_{K^*} = 898.6$ MeV, $\Gamma_{K^*}= 51.1$ MeV). This confirms that the scalar strange resonance can not contribute in a sizably way to the $D_{e4}$ observables.

\subsection{Determination of $|V_{cs}|$}

 In our fits we have fixed the CKM matrix element to its value quoted in Ref. \cite{PDG}, $|V_{cs}| = 0.975 \pm 0.006$. Since $A_1(0)$ is the global normalization constant of the decay distributions, we can translate its fitted values of Tables \ref{RF1} and \ref{RF1D0} into the following results (we assume isospin symmetry $A_1^{D^+\to K^{*0}}(0)=A_1^{D^0\to K^{*-}}(0)\equiv A_1(0)$) 
\begin{equation}
    |V_{cs}\cdot  A_1(0)| =\left\{\begin{array}{l} 0.611\pm 0.010, {\rm \ from\ D^+_{e4}} \\ 
    \\
      0.575\pm 0.012, {\rm \ from\ D^0_{e4}}\  \end{array} \right. 
\end{equation}
 These two results should be the same owing to isospin symmetry, however, their central values differ by 2.3 times their quoted errors. We can take their weighted average and inflate the error by the scale factor ($S=2.3$). We get 
  $$|V_{cs}\cdot  A_1(0)| = 0.596\pm 0.018 $$ 
  
   Now, using for instance the theoretical prediction for the axial weak charge $A_1(0)=0.619(17)$ shown in Table \ref{TableFFValues}  \cite{BESIII1}, we get 
\begin{equation}\label{vcs2}
|V_{cs}|=0.963\pm 0.041 
\end{equation}

We must keep in mind that this estimate is limited from the poor information we have on experimental data. Also, the theoretical error on $A_1(0)$ is not reported in the original calculation.

\begin{figure}[ht]
\centering
\includegraphics[width=9.6cm]{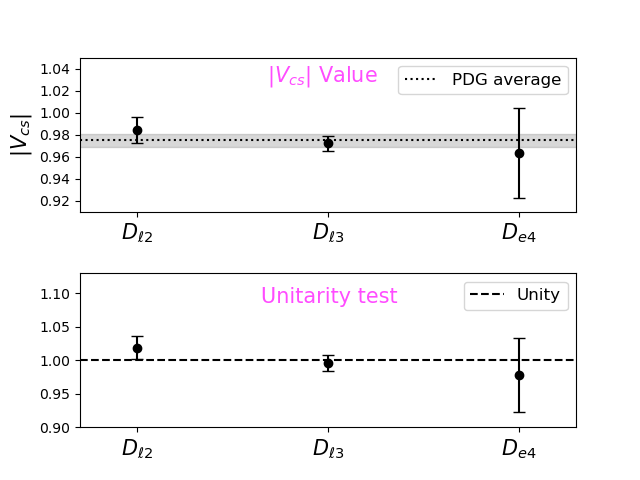}
  \caption{Upper panel: $|V_{cs}|$ values obtained, respectively, from the $D_{\ell 2}$, $D_{\ell 3}$ (see \cite{PDG} and references therein), and the one extracted in this work using $D_{e4}$ decays. The shaded gray band shows the PDG average value of $|V_{cs}|$. Lower panel: the corresponding test of the CKM unitarity  using the $|V_{cd}|$ and $|V_{cb}|$ values from Ref. \cite{PDG}. 
  }
  \label{Vcs_Uni}
\end{figure}

In Figure \ref{Vcs_Uni} we compare Eq. (\eqref{vcs2}) with the determinations of this parameter $|V_{cs}|=0.984\pm 0.012$ from leptonic $D_s\to \ell^+ \nu_{\ell}$ ($D_{s\ \ell 2}$) and $|V_{cs}|=0.972\pm 0.007$ from semileptonic $D\to Ke^+\nu_e$ ($D_{e3}$) decays \cite{PDG}. Currently, the precision of $|V_{cs}|$ obtained from $D_{e4}$ decays currently is not competitive compared with determinations from other sources. Further theoretical determinations of the form factors at zero momentum transfers together with improvements in experimental data, would be useful to reduce the error of $|V_{cs}|$ from $D_{e4}$ decays, providing eventually a competitive determination of this parameter. 

\section{Conclusions} \label{Conclusions}

We have analyzed the four-body semileptonic decays $D_{e4}$  within a model where the $K\pi$ system is dominated by the $K^*(892)$ resonance and a novel contribution $D^*$ pole meson is also considered. We find that the $D^*$ pole contribution can mimic the effects in the decay rate provided by the $S$-wave background  $D_{e4}$ decays in the fits of Refs.\cite{BESIII1, BESIIID0}. More detailed analyses in the future with more data will help to elucidate and disentangle the dynamical contributions to the hadronic weak vertex. Meanwhile, we find very interesting that current data the  $D^*$ pole contribution could mimic the effect of the $S$-wave contribution in the branching fraction.

The fits exhibit a sizable correlation of the $D^*$-pole contribution with the form factors (
at $q^2=0$) of the dominant $D\to K^*$ transition, modifying also the normalization of the decay distributions at zero momentum transfer. From the latter, we can extract an average of the $|V_{cs}A_1(0)|$
 product by assuming isospin symmetry between $D^+_{e4}$ and $D^0_{e4}$ decays. By relying in the prediction of Ref. \cite{BESIII1} for the weak axial charge $A_1(0)$, we extract the value of $|V_{cs}|$ from $D_{e4}$ decays. The central value of $|V_{cs}|(D_{e4})$ turns out to be smaller than but consistent with other (most precise) determinations. Future improvements in theoretical calculations of the $D\to K^*$ and $D^* \to K$ form factors together with more precise measurements of the branching fractions and other observables of $D_{e4}$ decays at Belle II and BESIII experiments, will contribute to precise determinations of $|V_{cs}|$. 

 [Note added: After completing this work, we became aware of Ref. \cite{Bajc:1997nx}. Using a framework that combines heavy quark effective theory with chiral perturbation theory, the authors \cite{Bajc:1997nx} calculate the resonant and non-resonant contributions to the form factors of the $D^+_{e4}$ decay. The non-resonant contributions, identified with the S-wave, also contain the  $D^*$-pole. Their results indicate a non-resonant contribution of the order of $3-11\% $ to the branching fraction of $D^+_{e4}$ for different values of input parameters, consistent with experimental results and with the present work.]

\section*{Acknowledgments}
J.L.G.S. thanks Secretaría de Ciencia, Humanidades, Tecnología e Innovación (SECIHTI) for support through {\it Estancias Posdoctorales por México} and {\it SNII} programs. G.L.C. acknowledges support from SECIHTI programs Ciencia de Frontera (project No. CBF2023-2024-3226) and SNII. We are grateful to Fenfen An (BESIII collaboration) for sharing $D_{e4}^+$ decay data used in this work.

\appendix

\section{Four-body kinematics}\label{Appendix  A}
The kinematics of the generic four-body decay $D(p, M) \to K(P_1, m_1) \pi (P_2, m_2) \ell^+ (P_3, m_3) \nu_{\ell}(P_4, m_4)$, with four-momenta and the masses given within parenthesis, can be described in terms of five independent variables: ($s_{12}$, $s_{34}$, $\theta_{12}$, $\theta_{34}$ and $\phi$) \cite{Cabibbo}.  They are defined as: $s_{12} = P^2 = (P_{1} + P_{2})^2$, the invariant-mass of the pseudoscalar mesons system; $s_{34} = L^2 = (P_{3} + P_{4})^2$, the invariant-mass of the leptonic pair; $\theta_{12}$ ($\theta_{34}$), the angle between the $\pi$ meson (the neutrino) and the $D$ meson three-momenta in the pseudoscalar mesons (leptonic) system rest frame.
Finally, $\phi$ is the angle between the planes defined by pseudoscalar mesons and the leptons pairs. This set of variables is shown in Figure \ref{KinFig}.

 The integration limits for the angular variables are:  $ 0 \leq \theta_{12},\ \theta_{34} \leq \pi$ and $0\leq \phi \leq 2\pi$. The limits on the invariant masses depend upon the order of integration. If the last integration is carried out over $s_{12}$ we have:
\begin{eqnarray}
&& m_1+m_2 \leq \sqrt{s_{12}} \leq M-m_3-m_4\ , \\  
&& m_3+m_4 \leq \sqrt{s_{34}} \leq M-\sqrt{s_{12}}\ .
\end{eqnarray}
In the case that the last integration is carried over $s_{34}$, we have:
\begin{eqnarray}
&& m_3+m_4 \leq \sqrt{s_{34}} \leq M-m_1-m_2\ , \\  
&& m_1+m_2 \leq \sqrt{s_{12}} \leq M-\sqrt{s_{34}}\ .
\end{eqnarray}

In terms of this set of independent variables, the scalar products of the four momenta pairs ($P=P_1+P_2, \ Q=P_1-P_2\ L=P_3+P_4$ and $N=P_3-P_4$) can be expressed as follows
\begin{align*} 
P^2 &= s_{12}, & Q^2 &= 2(m_{1}^2+m_{2}^2)-s_{12}, \nonumber\\
L^2 &= s_{34}, & N^2 &= 2(m_{3}^2+m_{4}^2)-s_{34}, \nonumber\\
\end{align*}

\begin{eqnarray}\label{KIN}
P\cdot Q &=& m_{1}^2-m_{2}^2, \nonumber\\
L\cdot N &=& m_{3}^2 - m_{4}^2, \nonumber\\
P\cdot L &=& \frac{1}{2}(M^2-s_{12}-s_{34}), \nonumber\\
P\cdot N &=& b_{1} - c_{1}\cos\theta_{34}, \nonumber\\
Q\cdot L &=& P\cdot L \frac{P\cdot Q}{s_{12}} + X\beta_{12} \cos\theta_{12}, \nonumber\\Q\cdot N &=& b_{3}-c_{2}\cos \theta_{34}-d \sin\theta_{34}\cos\phi\ .
\end{eqnarray}
In the limit  $m_{3}=m_{4}=0$, these relations coincide with the ones given in  the reference \cite{Bijnens}.

The differential decay rate is given by

\begin{equation}
	d \Gamma = \frac{X \beta_{12} \beta_{34}}{4(4\pi)^6 M^3} \overline{\left | \mathcal{M} \right | ^2} \sin \theta_{12} \sin \theta_{34}ds_{12} ds_{34} d \theta_{12} d \theta_{34} d \phi \ ,
\end{equation}
where $\overline{|\mathcal{M}|^2}$ is the unpolarized decay probability.  The other quantities defined in this decay probability are
\begin{eqnarray}
X & = & \frac{1}{2} \lambda ^{1/2} (M^2, s_{12}, s_{34}), \nonumber \\
\beta_{12} & = & \frac{1}{s_{12}} \lambda ^{1/2}(s_{12}, m_{1}^2, m_{2}^2), \nonumber \\
\beta_{34} & = & \frac{1}{s_{34}} \lambda ^{1/2}(s_{34}, m_{3}^2, m_{4}^2) \ .\nonumber 
\end{eqnarray}

The probability $\overline{|\mathcal{M}|^2}$ can be expanded explicitly in terms of the angles $\theta_{34}$ and $\phi$, whose coefficients depend only on the remaining variables $s_{12}$, $s_{34}$, and $\theta_{12}$ (see Eq. (\ref{angdist}) below). Such a decomposition allows to integrate over $\theta_{34}$ and $\phi$ angles in the case that the decay probability does not depend on them \cite{Pais-Treiman, Semi-Bijnens}. This distribution is given in Appendix \ref{Appendix B}.

\section{Unpolarized Squared Amplitude}\label{Appendix B}
 The spin-averaged decay probability  $\overline{|\mathcal{M}|^2}$, where ${\cal M}$ was defined in Eqs. (\ref{amplitude}) and (\ref{most-general-decomposition}), can be written as follows:

\begin{strip}
\begin{eqnarray}\label{USA}
\overline{\left | \mathcal{M} \right | ^2} 
&=& 
\frac{4 \left | V_{qq'} \right | ^2 G_{F}^2}{ M^2} \left\{ \frac{\left | H \right | ^2}{M^4} \left[ - \frac{1}{2} \epsilon_{\mu \nu \rho \sigma}L^{\mu} N^{\nu} P^{\rho} Q^{\sigma} \epsilon_{\delta \alpha \beta \gamma} L^{\delta} N^{\alpha} P^{\beta} Q^{\gamma}  
- \frac{1}{4} (L^2 - N^2) \epsilon^{\delta}_{\nu \rho \sigma} L^{\nu} P^{\rho} Q^{\sigma} \epsilon_{\delta \alpha \beta \gamma} L^{\alpha} P^{\beta} Q^{\gamma} \right] \right. \nonumber \\
& \quad \quad + & 
\left. \frac{\left | F \right | ^2}{4} \left[ 2 (P \cdot L^2 - P \cdot N^2) - P^2 (L^2 - N^2) \right] 
+ 
\frac{\left | G \right | ^2}{4} \left[ 2 (Q \cdot L^2 - Q \cdot N^2) - Q^2 (L^2 - N^2)  \right] \right. \nonumber \\
& \quad \quad + & 
\left. \frac{\left | R \right | ^2}{4} \left[ L^2( L^2 + N^2) - 2 L\cdot N^2  \right] 
+ 
\frac{1}{2} \text{Re}(FG^{*}) \left[ 2 P \cdot L (Q \cdot L) - 2 P \cdot N (Q \cdot N) - P \cdot Q (L^2 - N^2) \right] \right. \nonumber \\
& \quad \quad + & 
\left. \frac{1}{2} \text{Re}(FR^{*}) \left[ P \cdot L (L^2 + N^2) - 2 L \cdot N (P \cdot N) \right]  + 
\frac{1}{2} \text{Re}(GR^{*}) \left[ Q \cdot L (L^2 + N^2) - 2 L \cdot N (Q \cdot N) \right] \right. \nonumber \\
 & \quad \quad + & 
 \left. \frac{1}{M^2} \left[ \text{Re}(HF^{*}) \epsilon^{\mu \lambda \delta \eta} N_{\lambda}  P_{\delta} L_{\eta}  + \text{Re}(HG^{*}) \epsilon^{\mu \lambda \delta \eta} N_{\lambda}  Q_{\delta} L_{\eta}\right] \epsilon_{\mu \nu \rho \sigma} L^{\nu} P^{\rho} Q^{\sigma} \right. \nonumber \\
& \quad \quad - &
 \left. \frac{1}{M^2} \left[ \text{Im}(HF^{*}) P \cdot N + \text{Im}(HG^{*}) Q\cdot N + \text{Im}(HR^{*}) L\cdot N \right] \epsilon_{\mu \nu \rho \sigma} L^{\mu} N^{\nu} P^{\rho} Q^{\sigma} \right. \nonumber \\
& \quad \quad + & 
\left. \text{Im}(FG^{*}) \epsilon_{\mu \nu \rho \sigma} L^{\mu} N^{\nu} P^{\rho} Q^{\sigma} \right\} 
\end{eqnarray}

This expression, quadratic in the form factors, is the most general one. We have not neglected the lepton masses in the final states, therefore it can be used to test lepton universality in the ratios of decay rates for $D_{\mu 4}/D_{e4}$. We have checked this expression by comparing it with the one derived in  \cite{Semi-Bijnens} for the case of $K_{\ell 4}$ decays.

\subsection{Angular distribution}
In order to perform the integrations more easily, we can write the squared unpolarized probability as an expansion in the angular variables $(\theta_{34}, \phi)$ \cite{Wise,Semi-Bijnens}. After integration over these two angular variables, we end up with a distribution on the remaining variables $s_{12}$, $s_{34}$ and $\theta_{12}$. Up to an overall constant, we can write  $\overline{|\mathcal{M}|^2}\sim I$, where $I$ can be written in terms of nine terms$I_{1,..,9}$:  

\begin{eqnarray} \label{angdist}
I &=& I_{1} + I_{2}\cos 2\theta_{34} + I_{3}\sin^2 \theta_{34}  +  
I_{4}\sin 2\theta_{34}\cos \phi + I_{5}\sin \theta_{34}\cos \phi \nonumber \\
& \quad \quad + & I_{6}\cos \theta_{34} + I_{7}\sin \theta_{34}\sin \phi + 
I_{8}\sin 2\theta_{34}\sin \phi + I_{9}\sin^2 \theta_{34}\sin 2\phi , 
\end{eqnarray}

\noindent where the $I_{1,2,...,9}$ are functions of the remaining variables: ($s_{12}$, $s_{34}$, $\theta_{12}$). The explicit expressions for these coefficients are  

\begin{eqnarray}\label{Large_Is}
I_{1} &=& \frac{\left | H \right |^2}{4M^6}\left\{ -\frac{d^2}{2}X^2 - \left( L^2-N^2 \right)\left[ Q^2\left(P\cdot L\right)^2 -2P\cdot L \left( Q\cdot L \right) P\cdot Q + L^2\left( P\cdot Q \right)^2 \right. \right. \nonumber \\
 &\quad \quad + & \left.\left.
 P^2\left( Q\cdot L \right)^2 - P^2L^2Q^2\right] \right\} 
+ \frac{\left | R \right |^2}{4M^2}\left[  L^2\left( L^2+N^2 \right)-2\left(L\cdot N\right)^2 \right] \nonumber \\
&\quad \quad + &
\frac{\left | F \right |^2}{4M^2}\left\{ 2\left[ \left( P\cdot L \right)^2-b_{1}^2-\frac{c_{1}^2}{2} \right] -P^2\left( L^2-N^2 \right) \right\} \nonumber \\
&\quad \quad + &
\frac{\left | G \right |^2}{4M^2} \left\{ 2\left[ \left( Q\cdot L \right)^2 -b_{3}^2-\frac{c_{2}^2}{2} -\frac{d^2}{4} \right] -Q^2\left( L^2-N^2 \right)\right\}\nonumber \\
& \quad \quad + &
\frac{\text{Re}\left( FG^{*} \right)}{2M^2}\left[ 2P\cdot L \left( Q\cdot L \right) - 2b_{1}b_{3} - c_{1}c_{2} - P\cdot Q \left( L^2-N^2 \right) \right]\nonumber \\
&\quad \quad + & 
\frac{\text{Re}\left( FR^{*} \right)}{2M^2}\left[ P\cdot L \left( L^2+N^2 \right) - 2b_{1}L\cdot N \right] \nonumber\\
&\quad \quad + &
 \frac{\text{Re}\left( GR^{*} \right)}{2M^2}\left[ Q\cdot L \left( L^2+N^2 \right) -2b_{3}L\cdot N\right] \nonumber \\
&\quad \quad + &
\frac{\text{Re}\left( HF^{*} \right)}{M^4} \left\{ \left[ P^2L^2 - \left( P\cdot L \right)^2 \right]b_{3} + \left[ P\cdot L\left( Q\cdot L \right) - \left( P\cdot Q \right) L^2 \right]b_{1} \right. \nonumber \\
&\quad \quad + & \left.
 L\cdot N \left[ P\cdot Q \left( P\cdot L\right) - P^2\left( Q\cdot L \right) \right] \right\}\nonumber \\
&\quad \quad + &
\frac{\text{Re}\left( HG^{*} \right)}{M^4} \left\{ \left[ \left( Q\cdot L \right)^2 - Q^2L^2\right]b_{1} + \left[ \left(P\cdot Q\right) L^2 - P\cdot L \left( Q\cdot L \right) \right]b_{3} \right. \nonumber \\
&\quad \quad + & \left. 
L\cdot N \left[ Q^2 \left( P\cdot L\right) - P\cdot Q\left( Q\cdot L \right) \right] \right\}, \nonumber \\
I_{2}&=&\frac{1}{2M^2}\left[ \frac{\left | H \right |^2 }{4M^4}X^2d^2 - \frac{\left | F \right |^2 }{2}c_{1}^2 - \frac{\left | G \right |^2 }{2}\left( c_{2}^2-\frac{d^2}{2} \right) - \text{Re}\left( FG^{*} \right)c_{1}c_{2}  \right],  \nonumber \\
I_{3}&=& \frac{d^2}{4M^2}\left[ \frac{\left | H \right |^2}{M^4}X^2 - \left | G \right |^2 \right], \nonumber \\
I_{4}&=& -\frac{1}{2M^2}\left[ \left | G \right |^2 c_{2}d  - \text{Re}\left( FG^{*} \right)c_{1}d \right], \nonumber \\
I_{5} &=& \frac{d}{M^2}\left\{ \left | G \right |^2b_{3} + \text{Re}\left( FG^{*} \right)b_{1} + \text{Re}\left( GR^{*} \right)L\cdot N \right. \nonumber \\
&\quad \quad + & \left. 
\frac{\text{Re}\left( HF^{*} \right)}{M^2} X^2 - \frac{\text{Re}\left( HG^{*} \right)}{M^2}\left[ \left(P\cdot Q\right)L^2 - P\cdot L \left( Q\cdot L \right) \right] \right\},  \nonumber \\
I_{6} &=& \frac{1}{M^2}\left\{ \left | F \right |^2 b_{1}c_{1} +  \left | G \right |^2 b_{3}c_{2} + \text{Re}\left( FG^{*} \right)\left( b_{1}c_{2}-b_{3}c_{1} \right) \right. \nonumber \\
& \quad \quad + & \left.
\text{Re}\left( FR^{*} \right) L\cdot N c_{1} + \text{Re}\left( GR^{*} \right) L\cdot N c_{2} \right. \nonumber \\
& \quad \quad + & \left.
\frac{\text{Re}\left( HF^{*} \right)}{M^2}\left[ c_{2}X^2-c_{1}\left( P\cdot L \left( Q\cdot L - P\cdot Q L^2 \right) \right) \right] \right. \nonumber\\
& \quad \quad + & \left.
\frac{\text{Re}\left( HG^{*} \right)}{M^2}\left[ c_{1}\left(Q^2L^2 - \left( Q\cdot L \right)^2  \right) -c_{2}\left( \left(P\cdot Q\right)L^2 - P\cdot L \left( Q\cdot L \right) \right) \right] \right\}, \nonumber \\
I_{7} &=& \frac{Xd}{M^2}\left[ -\text{Im}\left( FG^{*} \right)  + \frac{\text{Im}\left( HF^{*} \right)}{M^2} b_{1} + \frac{\text{Im}\left( HG^{*} \right)}{M^2} b_{3} + \frac{\text{Im}\left( HR^{*} \right)}{M^2} L\cdot N\right], \nonumber \\
I_{8} &=& -\frac{Xd}{2M^4}\left[ \text{Im}\left( HF^{*} \right)c_{1} + \text{Im}\left( HG^{*} \right)c_{2} \right], \nonumber \\
I_{9} &=& -\frac{\text{Im}\left( HG^{*} \right)}{2M^4}Xd^2\ . 
\end{eqnarray}
\end{strip}

The quantities $b_1,  \ b_3,  \ c_2$ and $d$ are given by 

\begin{eqnarray}
b_{1} &=& \frac{ (P\cdot L) ( L\cdot N)}{s_{34}}, \nonumber \\
c_{1} &=& X\beta_{34} , \nonumber\\
b_{3} &=& \frac{(P\cdot L )( P\cdot Q)}{s_{12}}+\frac{L\cdot N}{s_{34}}X\beta_{12}\cos \theta_{12}, \nonumber\\
c_{2} &=& \frac{P\cdot Q}{s_{12}}X\beta_{34}+\beta_{12}\beta_{34}P\cdot L \cos \theta_{12}, \nonumber\\
d &=&\beta_{12}\beta_{34}(s_{12}s_{34})^{1/2}\sin\theta_{12} \ . \nonumber
\end{eqnarray}

\bibliographystyle{spphys}
\bibliography{bibfile}

\end{document}